\documentclass[aps,onecolumn,showpacs,groupedaddress,longbibliography,pre,10pt]{revtex4-1}

\usepackage{graphicx}
\usepackage{amsfonts}
\usepackage{amssymb}
\usepackage{amsmath}
\usepackage{xcolor}
\usepackage{soul} 
\usepackage{ulem}
\usepackage{enumitem}

\newcommand{\beq}{\begin{equation}}
\newcommand{\eeq}{\end{equation}}
\newcommand{\bea}{\begin{eqnarray}}
\newcommand{\eea}{\end{eqnarray}}

\begin{document}

\title{Unstable cracks trigger asymptotic rupture modes in bimaterial friction}

\author{H. Shlomai\textsuperscript{1}}
\author{D. S. Kammer\textsuperscript{2}}
\author{M. Adda-Bedia\textsuperscript{1,3}}
\author{R. Arias\textsuperscript{4}} 
\author{J. Fineberg\textsuperscript{1}}
\affiliation{\textsuperscript{1}The Racah Institute of Physics, The Hebrew University of Jerusalem, Jerusalem, Israel 91904}
\affiliation{\textsuperscript{2}Institute for Building Materials, ETH Z\"urich,  8093 Z\"urich, Switzerland}
\affiliation{\textsuperscript{3}Universit\'e de Lyon, Ecole Normale Sup\'erieure de Lyon, Universit\'e Claude Bernard, CNRS, Laboratoire de Physique, F-69342 Lyon, France}
\affiliation{\textsuperscript{4}Departamento de F\'{\i}sica, Facultad de Ciencias F\'{\i}sicas y Matem\'{a}ticas, Universidad de Chile, Santiago 8370449, Chile}

\begin{abstract}

The rupture of the interface joining two materials under frictional contact controls their macroscopic sliding. Interface rupture dynamics depend markedly on the mechanical properties of the bulk materials that bound the frictional interface. When the materials are similar, recent experimental and theoretical work has shown that shear cracks described by Linear Elastic Fracture Mechanics (LEFM) quantitatively describe the rupture of frictional interfaces. When the elastic properties of the two materials are dissimilar, many new effects take place that result from bimaterial coupling: the normal stress at the interface is elastodynamically coupled to local slip rates. At low rupture velocities, bimaterial coupling is not very significant and interface rupture is governed by `bimaterial cracks' that are described well by LEFM.  As rupture velocities increase, we  experimentally and theoretically show how bimaterial cracks become unstable at a subsonic critical rupture velocity, $c_T$. When the rupture direction opposes the direction of applied shear in the softer material, we show that $c_T$ is the  subsonic limiting velocity. When ruptures propagate in the direction of applied shear in the softer material, we demonstrate that $c_T$ provides an explanation for how and when slip pulses (new rupture modes characterized by spatially localized slip) are generated. This work completes the fundamental physical description of how the frictional rupture of bimaterial interfaces takes place. 

\end{abstract}

\date{\today}
\maketitle

\section{Introduction}
\label{sec:intro}

The traditional view of friction has been to consider only the centers of mass of sliding bodies in frictional contact. Starting from the work of  Da~Vinci~\cite{bowden2001friction}, Amontons and Coulomb~\cite{Deresiewicz1988} and later work by Dieterich~\cite{dieterich1979modeling} and Ruina~\cite{ruina1983slip}, this view suggests that a characteristic ``friction coefficient" is sufficient to describe the onset of frictional motion.  Recent experiments have, however, highlighted the importance of considering the spatial degrees of freedom within the frictional interface that separates two contacting bodies ~\cite{rubinstein2004detachment,ben2010dynamics,Ben-David2010Nature,Ben-David2011,Svetlizky2014,shlomai2016structure,Bayart2016a,Svetlizky2017a,Bayart2019,Svetlizky2020,shlomai_PNAS,shlomai_jgr,rosakis2002intersonic,ohnaka1999scaling,Xu_Fukuyama_2018,kammer2014existence}. This work has demonstrated that the onset of frictional sliding is mediated by rupture fronts that propagate along the interface, while detaching the contacts that form it.  When the contacting bodies are identical (`homogeneous interfaces'), experiments have, moreover, demonstrated~\cite{Svetlizky2014} that the singular form and dynamics of these fronts can be quantitatively described by solutions for shear cracks~\cite{freund1998dynamic} that have been derived in the framework of linear elastic fracture mechanics (LEFM). While there are still questions regarding the validity of this approach for all classes of friction laws~\cite{Barras_2020}, these experiments supported early theoretical suggestions~\cite{palmer1973growth,das2003dynamic} that earthquakes may be described as shear cracks. 
 
Here, we consider `bimaterial interfaces'; frictional interfaces that are bounded by materials having either different elastic properties or different geometrical shapes~\cite{AldamBouchbinder2016GeometricBimaterial}. Bimaterial interfaces are the most general type of frictional interfaces. Early work by Weertman~\cite{weertman1963dislocations,weertman1980unstable} suggested that the dynamics of these interfaces have unique properties; the loss of symmetry across the interface introduces coupling between their differential motion (slip) and the normal stresses at the interface. Moreover, breaking the up-down symmetry across the interface induces either normal stress reduction or enhancement with slip, depending on a rupture's lateral propagation direction. Numerical calculations have predicted that this coupling gives rise~\cite{AmpueroBenZion2008,Scala2017} to both directional effects as well as to interesting modes of rupture called `slip pulses' in which slip is highly localized at the rupture tip. We note that slip pulses are not solely generated by bimaterial coupling, they have also been numerically observed under strong velocity weakening of friction~\cite{perrin1995self} that can be induced, for example, by poroeleastic~\cite{Heimisson2019} or thermal pressurization~\cite{brantut2019stability} effects.

Slip pulses in bimaterials have been observed experimentally~\cite{xia2005laboratory,lykotrafitis2006self,shlomai2016structure}. Recent work~\cite{shlomai_PNAS} has further demonstrated that bimaterial ruptures start as `bimaterial cracks', or crack-like rupture fronts. These are described by solutions that are analogous to the singular LEFM shear crack solutions along homogeneous interfaces, when the difference in the elastic properties of the materials forming the interface is accounted for.  Slip pulses are formed only after bimaterial cracks accelerate to very high velocities~\cite{shlomai_PNAS}, where bimaterial cracks appear to lose stability. Neither the stability of bimaterial cracks nor the transition to slip pulses within frictional interfaces is understood. 

What are the characteristic velocity scales for the bimaterial problem? In the subsonic fracture of homogeneous materials, the propagation speed of either tensile or shear cracks is limited by the Rayleigh wave speed~\cite{rayleigh1885waves}, $c_R$. Stroh introduced a kinematic argument to explain the origin of this limiting speed~\cite{stroh1957theory}. Were a crack to require no energy to break bonds, then it could be considered to be a disturbance propagating along the free surface (or a surface wave) that separates the two half-spaces above and below the fracture plane. $c_R$ is the limiting speed for a crack simply because, except for $c_R$, the dispersion relation for surface waves has no solution. An energy argument produces the same result; $c_R$ is the limiting propagation speed above which the energy flow to a moving crack tip (the energy release rate) becomes negative - an unphysical condition. The Stroh approach has been applied to bimaterial interfaces to define the concept of a `generalized' Rayleigh wave speed, $c_{GR}$, that corresponds to the velocity of nondispersive interfacial waves in frictionless contact between dissimilar materials~\cite{stoneley1924elastic,weertman1963dislocations,gol1967surface}. For moderate ratios of the elastic moduli (material contrasts) of the two bounding materials, $c_{GR}$ exists and its value lies between the $c_R$ of both materials. Otherwise, $c_{GR}$ is not defined. For material combinations where $c_{GR}$ exists, it was established~\cite{adams1995self,ranjith2001slip,AldamBouchbinder2016GeometricBimaterial} that steady frictional sliding along a bimaterial interface with Coulomb friction acting at the interface is ill-posed in the sense that interfacial disturbances of all wavelengths are unstable. Nonetheless, Weertman~\cite{weertman1980unstable} argued that when $c_{GR}$ exists, a self-healing slip pulse can propagate along the frictional interface at a velocity that is precisely $c_{GR}$, even when the remote shear stress is smaller than the frictional strength of the interface.

The fact that $c_{GR}$ is not defined for all bimaterial contrasts suggests that it is not the only relevant velocity scale to be considered in bimaterial rupture. There have been numerous theoretical studies of dynamic interfacial fracture of bimaterials~\cite{gol1967surface,willis1971fracture,atkinson1977dynamic,yang1991mechanics} that investigated the limiting behavior of the crack tip and led to different conclusions. Among these, the following claims regarding the terminal speed have been made: it is the lower Rayleigh wave speed of both materials (that corresponds to the $c_R$ of the softer material)~\cite{atkinson1977dynamic}, it is slightly larger than this characteristic velocity~\cite{willis1971fracture}, or, even, that it does not exist at all~\cite{yang1991mechanics}. From the experimental side, shear fracture was precipitated by impacting a bimaterial system composed of different bulk blocks connected by weakly bonded interface~\cite{liu1993highly,lambros1995shear}. These studies revealed that interfacial crack tip speeds in bimaterials can exceed not only the $c_R$ of the softer material but also its shear wave speed. Such `supershear' modes were also later observed~\cite{XiaRosakis2005SupershearTransition,shlomai2016structure,shlomai_jgr} in frictional bimaterial experiments. While certainly related to bimaterial frictional ruptures, it is still unclear  how the two systems map to one another. One important difference is that bimaterial frictional ruptures retain contact between materials along the crack faces, whereas the fracture surface of bimaterial  interfacial cracks is assumed to be  stress free. Interestingly, the asymptotic fields of each type of crack behave differently near the tip; interfacial cracks appear to involve complex fields~\cite{Rice1988Elastic,yang1991mechanics} while near-tip fields are real for frictional cracks~\cite{shlomai_PNAS}.

In this work we take a closer look at the stability of subsonic frictional ruptures in  bimaterial friction.  We first derive and examine the asymptotic behavior of subsonic frictional crack solutions. We reveal that solutions for the most physical friction law yield the same solutions as bimaterial shear cracks, with near-tip stresses found to be always square root singular and real. This behavior enables us to harness energy considerations to investigate the local dissipation at the crack tip as for classical fracture. We find that the existence of bimaterial contrast imposes significant asymmetry on the energy flux to the crack's tip. This asymmetry divides the energy release rate into two parts, each corresponding to the different half spaces formed by the interface. Careful analysis reveals a non-trivial energy flow that is qualitatively different than that of similar materials. This analysis enables us to associate a terminal rupture velocity to the positivity of both contributions. In contrast to $c_{GR}$, this new terminal velocity is always defined, whatever the material contrast. With these predictions in hand, we carefully analyze both new experimental results and numerical data. Our conclusion is that one should be careful to associate instabilities of propagating fronts to the more simple perturbative analysis, that does not take into account the existence of propagating rupture fronts. Moreover, our findings enable us to draw a full phase diagram that relates shear crack and slip pulses in both propagation directions.

The present approach significantly extends the results of our analysis in the transonic and supersonic regimes~\cite{shlomai_PNAS,shlomai_jgr} to the subsonic regime where all ruptures initiate. This work will allow for a more complete picture of the onset of frictional sliding between dissimilar materials and reveals a qualitatively new class of instabilities. 

\begin{figure}[tb]
\begin{center}
\includegraphics[width=0.4\linewidth]{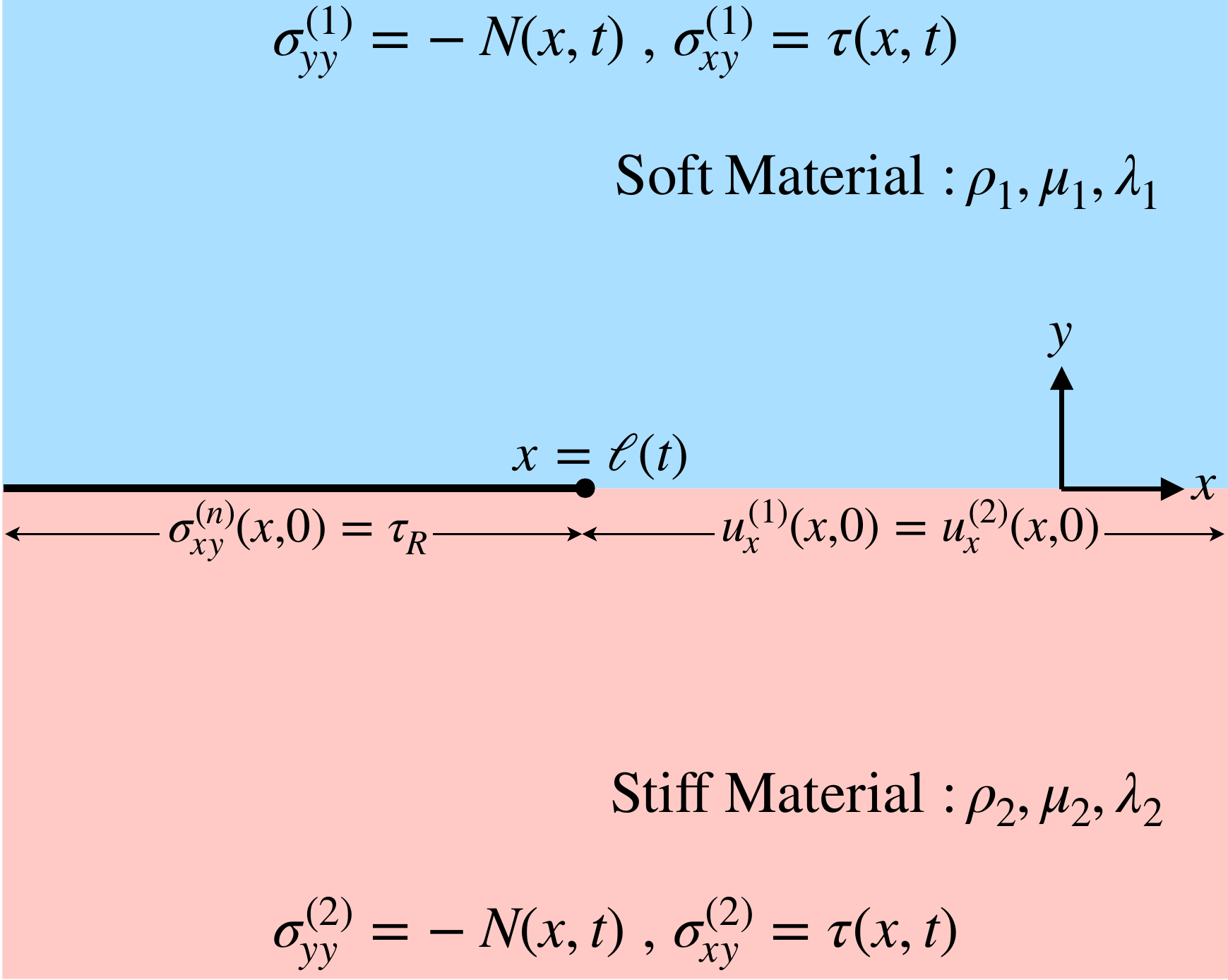}
\caption{Schematic of a dynamical frictional rupture propagating along a bimaterial interface. The material parameters and the main loading conditions are depicted.}
\label{fig:problem1}
\end{center}
\end{figure}

\section{Theory of bimaterial frictional rupture}
\label{sec:theory}
 
We start by presenting the general problem of a subsonic frictional rupture along a bimaterial interface as depicted in Fig.~\ref{fig:problem1}. Recently, this configuration has been used to determine the asymptotic behavior of the stress field near the rupture front in subsonic, transonic and supershear propagation regimes~\cite{shlomai_PNAS,shlomai_jgr}. These studies enabled successful theoretical characterization of various features observed in experiments. In the following, we set the framework for the present study and define the different associated physical quantities.

We consider a model problem of the dynamic deformation of a two-dimensional medium that results from shear rupture along a frictional interface. The interface is located on the plane $y=0$ that separates linear isotropic elastic half-spaces (see Fig.~\ref{fig:problem1}). The fields' motion and rupture propagation are in the $x$-direction and all variables are functions of $x$, $y$ and $t$. Shear and dilatational wave velocities are $c_{sn}=\sqrt{\mu_n/\rho_n}$ and $c_{dn}=\sqrt{(\lambda_n+2\mu_n)/\rho_n}$, where $\rho_n$ are mass densities, $\lambda_n$ and $\mu_n$ are Lam\'e coefficients, and the subscripts $n=1,2$ denote the top ($y>0$) and bottom ($y<0$) materials, respectively. Without loss of generality, we choose $c_{s1}<c_{s2}$ meaning that the top (bottom) material is the softer (stiffer) one. The applied shear and compression at the remote boundaries are $\tau(x,t)$ and $N(x,t)$ respectively. The local balance of linear momentum in the bulk requires that the elastic fields satisfy:
\beq
\frac{\partial\sigma^{(n)}_{ij}}{\partial x_j}=\rho_n \frac{\partial^2 u^{(n)}_i}{\partial t^2}\;,
\label{eq:bulk}
\eeq
where $\sigma^{(n)}_{ij}(x,y,t)$ and $u^{(n)}_i(x,y,t)$ are the elastic stress and displacement fields, respectively. In the following, the subscripts $(i,j)$ are two dimensional indices labeling the $x$ and $y$ directions. Along the bimaterial interface, a semi-infinite frictional in-plane rupture is located at $x_{tip}<\ell(t)$. The rupture front is propagating at a subsonic velocity $c(t)=\dot{\ell}(t)<c_{s1}$. Notice that, due to the asymmetry induced by material mismatch, rupture propagation in the positive and negative $x$-directions are not equivalent. For a frictional interface between dissimilar materials, propagation in the so-called positive (resp. negative) direction is characterized by a rupture front that is moving in the same (resp. opposite) direction as the slip direction of the softer material. In the convention of Fig.~\ref{fig:problem1}, propagation in the positive (resp. negative) direction corresponds to a loading configuration for which $\tau(x,t)>0$ (resp. $\tau(x,t)<0$).

Motivated by the experimental evidence of the persistence of crack face contact after the passage of a subsonic rupture front, we restrict our study to frictional contact in which both the traction and normal displacements are continuous across the entire bimaterial interface. Therefore, along the whole interface, the following boundary conditions are satisfied:
\begin{eqnarray}
 u^{(1)}_{y}( x, 0^+,t)&=& u^{(2)}_{y}( x, 0^-,t) \;,\label{ubc} \\
\sigma^{(1)}_{yy}( x, 0^+,t)&=&\sigma^{(2)}_{yy}( x, 0^-,t)\;,\\
\sigma^{(1)}_{xy}( x, 0^+,t)&=&\sigma^{(2)}_{xy}( x, 0^-,t)\;. \label{sbc} 
\end{eqnarray}
In order to solve the frictional rupture problem, still one should prescribe the slip conditions and/or the friction law along the interface. Ahead of the rupture front where no slip occurs one has
\begin{equation}
\delta(x,t)\equiv u^{(1)}_{x}(x,0^+,t)-u^{(2)}_{x}(x,0^-,t)=0\,;\qquad x>\ell(t) \;,
\label{eq:bcdelta}
\end{equation}
where $\delta(x,t)$ is the local slip at the interface. Behind the rupture front, the contact between the bodies is \textit{partially} broken inducing a drop of the shear stress accompanied by interface slip. The physical description of this phenomenon is given by the so-called friction law, which constitutes the boundary condition along the sliding region. Various friction laws have been proposed in the literature. Among them, the most popular one is the generalized Coulomb friction law. This criterion relates the local frictional resistance to the local normal stress at the interface through a material dependent friction coefficient. In the following analytical treatment, we assume a simpler scenario where sliding induces solely a drop of the shear stress to a constant residual value $\tau_R$:
 \begin{equation}
\sigma_{xy}^{(n)}(x,0,t)=\tau_R\,;\qquad x<\ell(t)\;.
\label{eq:bc-tau}
\end{equation}
Because this boundary condition does not induce coupling between normal and shear stresses, this type of frictional behavior is sometimes associated to a frictionless shear crack. While Eq.~(\ref{eq:bc-tau}) is the simplest friction law that one can consider, it is relevant so long as the material behavior deviates from the predictions of this law solely in a zone around the rupture front that is too small to be detected on the scale of this linear elastic model. From this perspective, any more elaborate friction law that allows for a constant residual stress in an intermediate region behind the rupture front can be considered as a cohesive model for which Eq.~(\ref{eq:bc-tau}) reproduces the elastic fields at scales larger than the size of the process zone.

\subsection{Asymptotic fields near a moving frictional rupture front}
\label{sec:fields}

The elastodynamic problem given by Eqs.~(\ref{eq:bulk})-(\ref{eq:bc-tau}) is well-posed once the remote loading and the motion of the rupture front are prescribed. In the following, we are interested in the universal properties of the asymptotic elastic fields in the vicinity of the propagating front that are independent of its dynamics and applied remote loading. This problem was previously solved in~\cite{deng1993propagating,shlomai_PNAS}. It was shown that the asymptotic stress and strain fields exhibit a universal square root singularity for any subsonic rupture front dynamics. In Appendix~\ref{app:fields}, the explicit behavior of the stress field is reported and the real nature of the singularity in comparison to other interfacial crack problems is discussed. Specifically along the interface $y=0$, the square root singular term of the shear stress component is given by
\beq
\sigma_{xy}^{(n)} (x\rightarrow\ell(t),0,t) \approx  K(t)\,\frac{ H(x-\ell(t)) }{\sqrt{2\pi (x-\ell(t))}} \; ,
\label{eq:sxy}
\eeq
where $H(.)$ is the Heaviside function and $K(t)$ is the so-called dynamic stress intensity factor. Notice that the sign of $K(t)$ traces the propagation direction. In the configuration of Fig.~\ref{fig:problem1} where the softer material occupies the half-plane $y>0$ and the rupture front propagates in the positive $x$-direction, the positive propagation direction corresponds to the case $K(t)>0$ and vice-versa. Correspondingly, the asymptotic behavior of the normal stress component at $y=0$ is given by
\beq
\sigma_{yy}^{(n)} (x\rightarrow\ell(t),0,t) \approx  K(t)\,W(c) \,\frac{H(\ell(t)-x) }{\sqrt{2\pi (\ell(t)-x)}}\; ,
\label{eq:syy}
\eeq
where the function $W(c)$, which we will name the Weertman function, is given by~\cite{weertman1980unstable}
\beq
W(c) = \frac{(1+b_1^2-2a_1b_1)\mu_2D_2-(1+b_2^2-2a_2b_2)\mu_1D_1}{a_1(1-b_1^2)\mu_2D_2+a_2(1-b_2^2)\mu_1D_1} \ ,
\label{eq:weertman}
\eeq
with
\beq
a_n=\sqrt{1-\frac{c^2}{c_{dn}^2}}\;, \quad b_n=\sqrt{1-\frac{c^2}{c_{sn}^2}}\;,\quad D_n=4 a_nb_n-(1+b_n^2)^2\;.
\label{eq:speeds}
\eeq
Recall that $c=\dot{\ell}(t)$ is the instantaneous speed of the rupture front and that the Rayleigh wave speed $c_{Rn}$ of each material is the positive root of the equation $D_n(c)=0$. The Weertman function highlights induced bimaterial coupling between slip and normal stress at the interface. Its behavior as function of the rupture speed has been studied in detail and was related to instability mechanisms for slip-pulse nucleation~\cite{weertman1980unstable,cochard2000fault}. For a homogeneous interface, $W(c)=0$, reflecting uncoupling between normal stress and slip at the interface. Moreover, the denominator of the Weertman function can have a real root for $c<c_{s1}$, which defines the generalized Rayleigh wave speed $c_{GR}$. This speed is only defined up to moderate bimaterial mismatches, $c_{GR}$ does not exist otherwise. Finally, when $c_{GR}$ exists $W(c)>0$ (resp. $W(c)<0$) for $c< c_{GR}$ (resp. $c_{GR}<c<c_{s1}$) and when $c_{GR}$ does not exist $W(c)>0$ for all $c<c_{s1}$. Note that when  $W(c)>0$, the sign of $K(t)$ (hence the propagation direction) determines whether the normal stress is enhanced or reduced with the slip (see Eq. \ref{eq:syy}).

\subsection{Energy concepts in dynamic frictional rupture}
\label{sec:energy}

A step forward in the characterization of the dynamics of frictional rupture consists in coupling the behavior of the elastic fields near the moving front to energy budgeting. For this purpose, one should compute the instantaneous rate of energy flow towards the frictional interface~\cite{freund1998dynamic}. The energy rate balance should, however, be performed carefully for the current case because of two features specific to the bimaterial frictional problem. First, the energy flux integral in the presence of compressive loading differs from the classical crack problem, even if the asymptotic elastic fields near the rupture tip behave similarly~\cite{palmer1973growth,rice1980mechanics}. This difference occurs for both homogeneous and bimaterial frictional interfaces. The second feature, which is specific to bimaterials, is that an interface separating two dissimilar bodies is a surface of discontinuity of both material mechanical properties and elastic fields. While the latter is common to all rupture modes, the former induces different balance equations of linear momentum in the top and bottom material, Eq.~(\ref{eq:bulk}). Therefore, the instantaneous energy rate balance needs to be computed for the volume of each material separately.

\begin{figure}[tb]
\begin{center}
\includegraphics[width=0.4\linewidth]{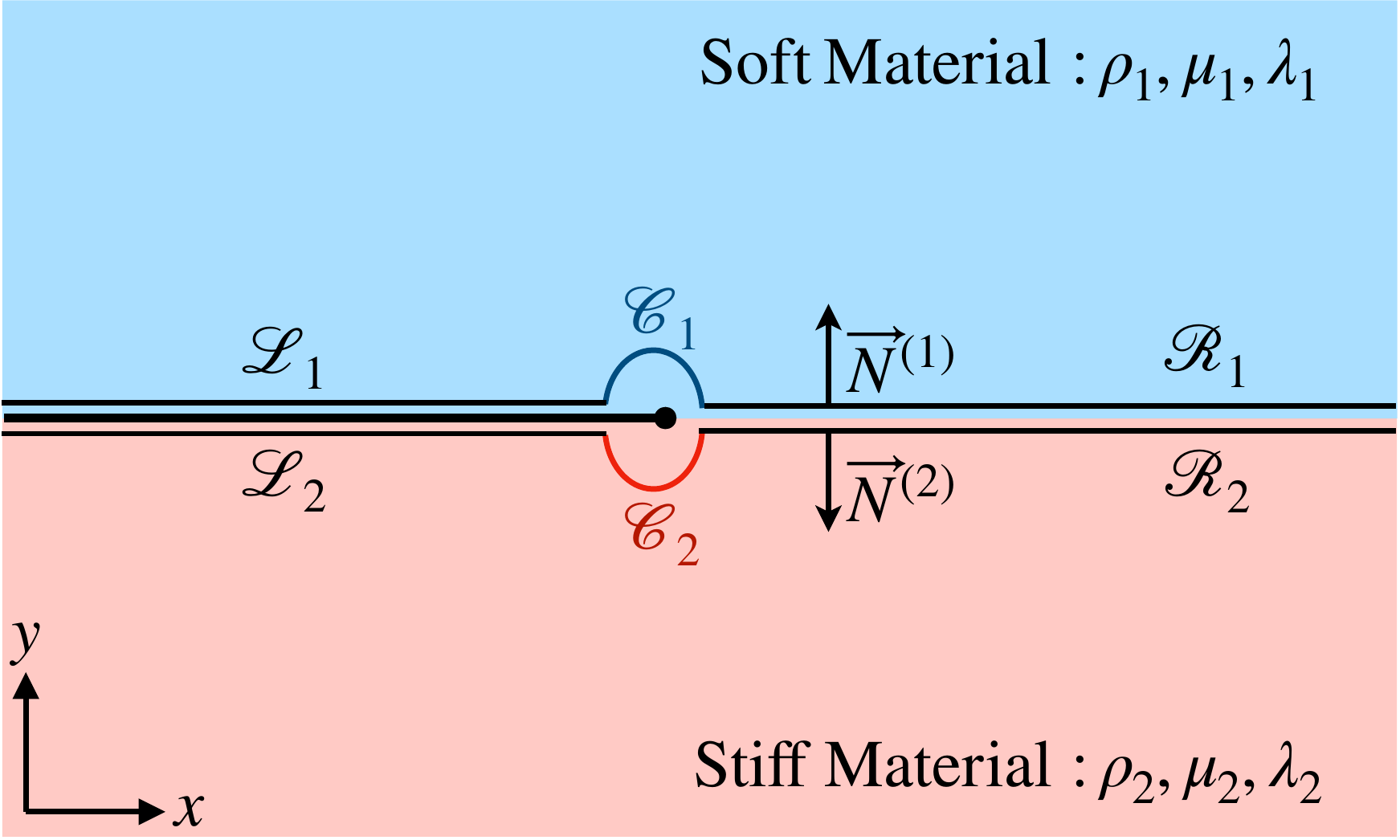}
\caption{Paths of integration involved in the computation of the rate of energy flow into the frictional interface. The paths ${\mathcal L}_n+{\mathcal C}_n+{\mathcal R}_n$ are open contours enveloping the bimaterial interface over which energy fluxes are exchanged. The net instantaneous rate of energy flow from the body towards the frictional interface is the sum of both contributions.}
\label{fig:problem2}
\end{center}
\end{figure}

Following the same approach as for a crack propagating in a homogenous medium~\cite{freund1998dynamic,adda1999dynamic,Adda1999}, one can show that the rate of mechanical energy flow, ${\mathcal G}^{(n)}$, out of each material and into the frictional interface is given by
\beq
{\mathcal G}^{(n)}=\frac{1}{c}\int_{{\mathcal L}_n+{\mathcal C}_n+{\mathcal R}_n}\left[\sigma^{(n)}_{ji}N_j^{(n)}\frac{\partial u^{(n)}_i}{\partial t}+\left(U^{(n)}+T^{(n)}\right)cN^{(n)}_x\right]dx\;,
\label{eq:defG}
\eeq
where, respectively for $n=1$ and $n=2$, the integral runs over and below the whole interface following the paths depicted in Fig.~\ref{fig:problem2}. Here, $T^{(n)}$ is the kinetic energy density and $U^{(n)}$ is the stress work density of each material, and $\vec{N}^{(n)}$ is a unit normal vector defined in Fig.~\ref{fig:problem2}. The total dynamic energy release rate, ${\mathcal G}$, resulting from the propagation of a frictional rupture is the sum ${\mathcal G}^{(1)}+{\mathcal G}^{(2)}$. Ahead of the rupture front, the continuity of both traction and displacement distributions across the interface leads to the integrals over ${\mathcal R}_n$ cancelling each other out. Note that while $\sigma_{xx}^{(n)}$ is discontinuous across a bimaterial interface, this stress component is not involved in the computation of ${\mathcal G}$ except, potentially, along the contour ${\mathcal C}_n$. Therefore, the energy release rate of a frictional rupture in the presence of compressive loading can be decomposed into two contributions~\cite{palmer1973growth,rice1980mechanics}
\beq
{\mathcal G}=G_{\mathrm{slip}}+G_{\mathrm{sep}}\;,
\label{eq:Gtot}
\eeq
with
\bea
G_{\mathrm{slip}}&=& \frac{1}{c}\int_{-\infty}^{\ell(t)}\tau_R\,\frac{\partial \delta}{\partial t}dx\;,
\label{eq:Gslide}\\
G_{\mathrm{sep}}&=&\sum_{n=1}^{2}G_n=\frac{1}{c}\sum_{n=1}^{2} \lim_{{\mathcal C}_n\rightarrow0}\int_{{\mathcal C}_n}\left[\sigma^{(n)}_{ji}N^{(n)}_j\frac{\partial u^{(n)}_i}{\partial t}+\left(U^{(n)}+T^{(n)}\right)cN^{(n)}_x\right]dx\;.
\label{eq:Gsep}
\eea
Eq.~(\ref{eq:Gslide}) results from the sum over $n$ of the integrals in Eq.~(\ref{eq:defG}) involving the paths ${\mathcal L}_n$ and the use of the boundary conditions~(\ref{ubc}-\ref{eq:bc-tau}). The physical quantity $G_{\mathrm{slip}}$ is the rate of energy provided by bulk materials along the sliding interface which should be dissipated because of the frictional process. On the other hand, Eq.~(\ref{eq:Gsep}) defines the mechanical energy released per unit crack advance; $G_{\mathrm{sep}}$ is the rate of energy provided locally at the rupture front to induce motion and should balance the resistance of bimaterial interface to rupture propagation. Notice that $G_{\mathrm{sep}}=G_1+G_2$ where $G_1$ (resp. $G_2$) is the rate of mechanical energy provided by the soft (resp. stiff) bulk material to the moving rupture front. The decomposition~(\ref{eq:Gtot}) of the total energy release rate into distinct contributions is relevant so long as there exists a separation of scales between the dissipation involved in material separation (`breaking' contacts) and frictional dissipation due to sliding. This behavior is closely related to how the frictional process is modeled. The assumptions that contact at the bimaterial interface is preserved after the passage of the rupture front and that the friction law is given by Eq.~(\ref{eq:bc-tau}) allows for such decomposition. These conditions induce the square root singular behavior of the stress field in the vicinity of the moving rupture tip that has been observed experimentally~\cite{shlomai_PNAS}.

\begin{figure}[tb]
\begin{center}
\includegraphics[width=0.4\linewidth]{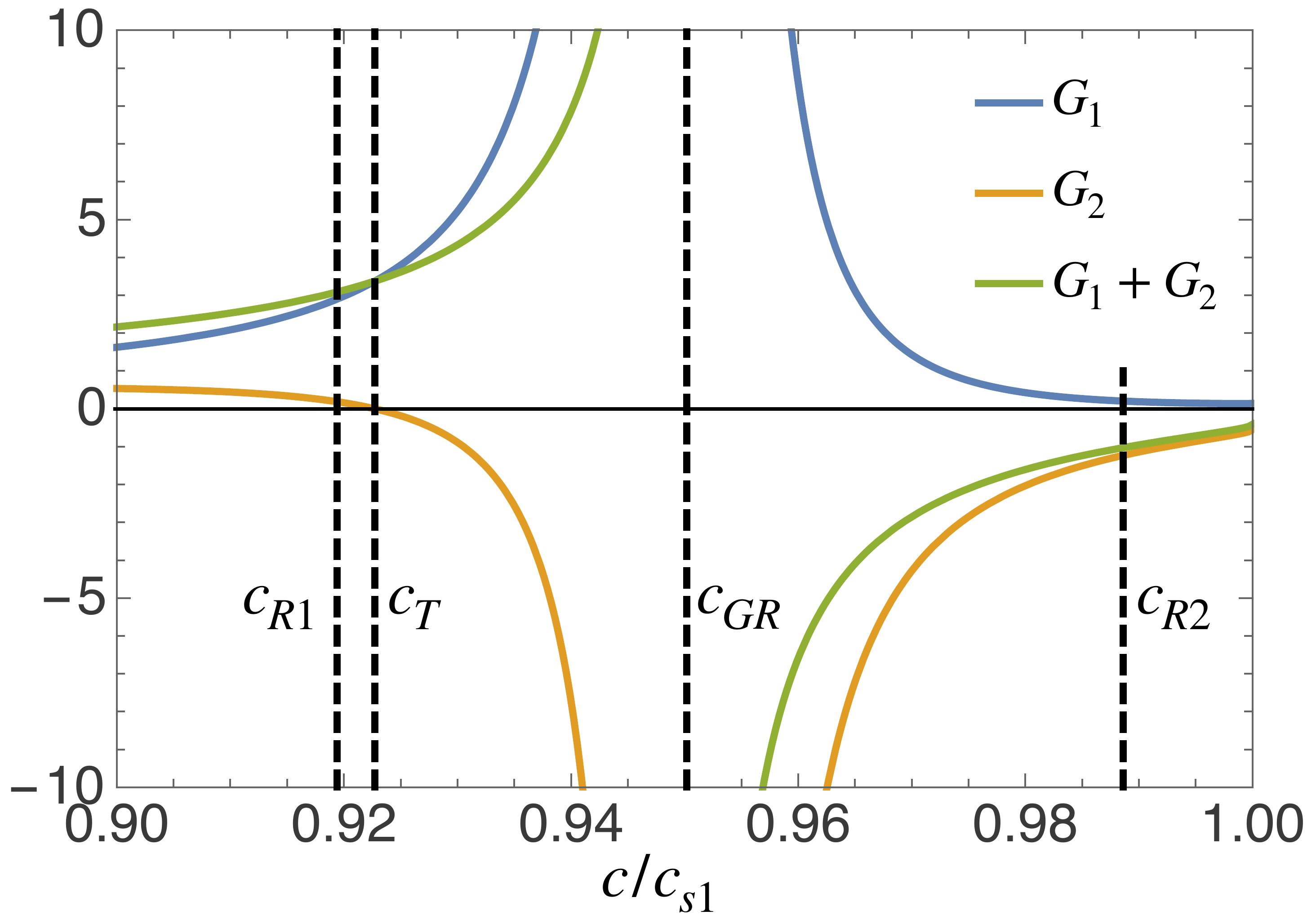}
\includegraphics[width=0.4\linewidth]{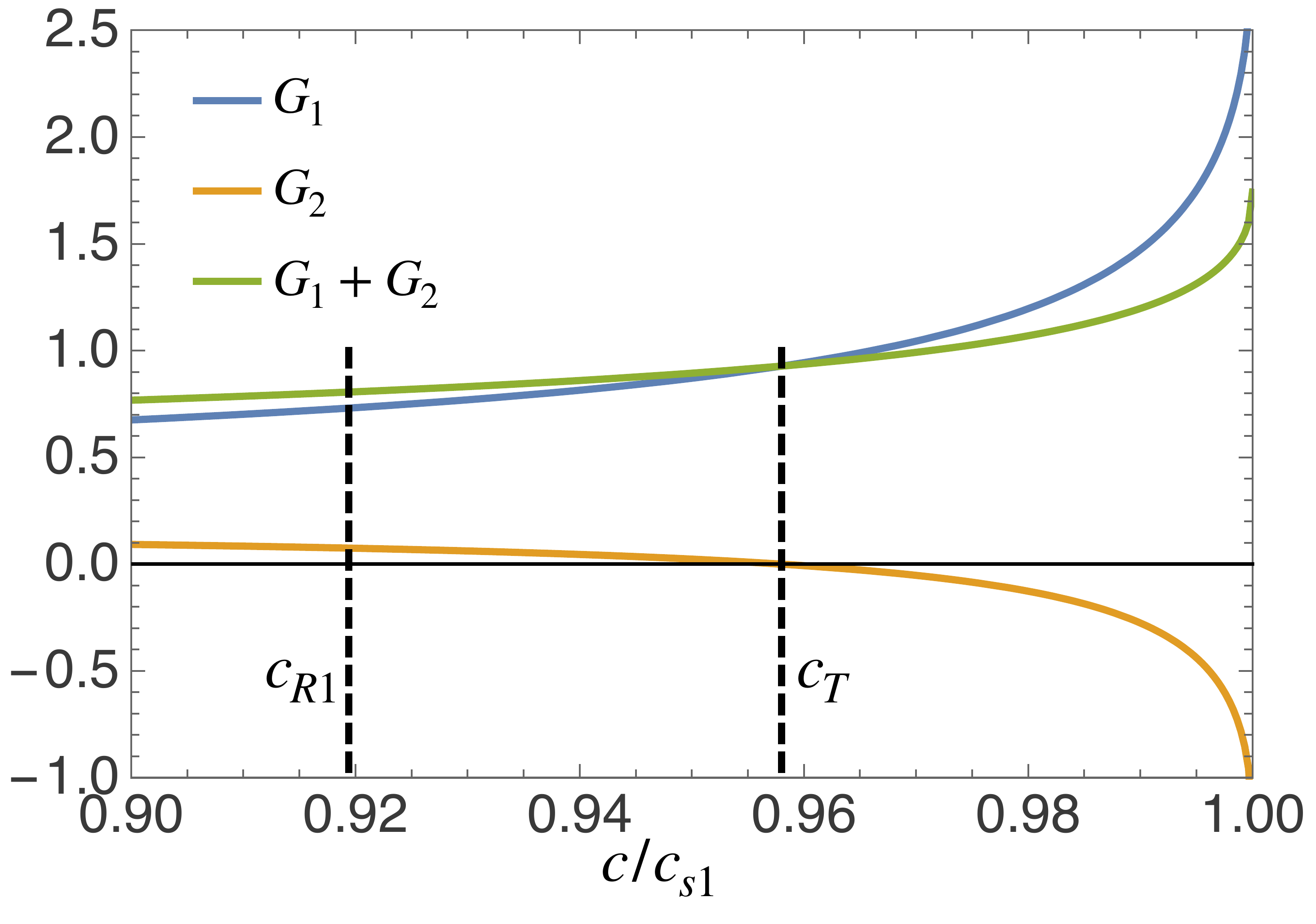}
\caption{The energy release rates $G_n$ and $G_{\mathrm{sep}}=G_1+G_2$ as functions of the rupture front speed $c$ in the region of interest $0.9c_{s1}\leq c\leq c_{s1}$ for two typical bimaterials. The bulk elastic parameters for both model bimaterials staisfy $\rho_1=\rho_2$, $c_{d1}/c_{s1}=c_{d2}/c_{s2}=\sqrt{3}$ and $c_{s1}/c_{s2}=\gamma$. Left: case $\gamma=0.93$ for which the generalized Rayleigh wave speed $c_{GR}$ exists. Right: case $\gamma=0.6$ for which $c_{GR}$ does not exist. In both figures, the energy release rates are non-dimensioned by $K^2/2\mu_1$ and the locations of the different characteristic velocities, especially the newly identified critical speed $c_T$, are indicated (see text).}
\label{fig:Gs}
\end{center}
\end{figure}

We now focus on the study of each of the {\it separate} energy release rates $G_n$. First, it can be shown that $G_n$ share the following properties with cracks propagating in homogeneous materials~\cite{freund1998dynamic,adda1999dynamic,Adda1999}: they are path independent and their computation involves the motion of rupture front only through its instantaneous crack tip speed $c(t)=\dot{\ell}(t)$. Since the interface separating two dissimilar bodies is, however, a surface of discontinuity of bulk material parameters, each contribution $G_n$ should be computed separately. In Appendix~\ref{app:ERR}, we show that
\bea
G_n&=&\frac{1}{4\mu_n } \left[1-\frac{a_n}{b_n}W^2(c)\right]\,A_{II}^{(n)}(c)\,K^2(t) \; ,
\label{eq:finalGn}
\eea
where, again, the Weertman function $W(c)$ is involved and
\beq
A_{II}^{(n)}(c)=\frac{(1-b_n^2)b_n}{D_n }\;,
\eeq
is the usual universal function that is involved in the energy release rate of mode~II crack propagation~\cite{freund1998dynamic,adda1999dynamic,Adda1999}. Eq.~(\ref{eq:finalGn}) is the main theoretical result of this study. It is important to stress that the functional forms of $G_n$ ($n=1,2$) given by Eq.~(\ref{eq:finalGn}) are universal, in the sense that they depend only on the instantaneous speed $c(t)=\dot{\ell}(t)$ (provided that $c(t)<c_{s1}$). As in homogeneous fracture dynamics, all information about loading conditions, propagation direction and bimaterial geometric scales are embedded in the dynamic stress intensity factor $K(t)$. Figure~\ref{fig:Gs} highlights the variation of the different energy release rates with rupture speed for two typical bimaterial mismatches. 

Below, we note the salient properties of $G_n$:
\begin{itemize}[leftmargin=*]

\item Equation~(\ref{eq:finalGn}) recovers the result~\cite{freund1998dynamic} of rupture propagating along a homogeneous interface separating two similar materials for which one has $W(c)=0$:
\beq
G_1^{(\mathrm{hom)}}=G_2^{(\mathrm{hom)}}=\frac{1}{2}G_{\mathrm{sep}}^{(\mathrm{hom)}}=\frac{1}{4\mu^{(\mathrm{hom)}}} A^{(\mathrm{hom)}}_{II}(c)\,K^2(t)\;.
\label{eq:Ghom}
\eeq
Note that the function $A^{(\mathrm{hom)}}_{II}(c)$ diverges at material's Rayleigh wave speed $c_R^{(\mathrm{hom)}}$, being positive for $c<c_R^{(\mathrm{hom)}}$ and negative for $c>c_R^{(\mathrm{hom)}}$.

\item In Fig.~\ref{fig:Gs}, the interest was focused on the variation of the different energy release rates for rupture speeds close to $c_{s1}$. Indeed, for small rupture speeds and irregardless of bimaterial mismatch, we find that the variations of $G_1$ ($G_2$) are moderate, positive, and monotonically increase (decrease) with $c$.

\item Figure~\ref{fig:Gs} shows that $G_n$ is regular at $c=c_{Rn}$, the Rayleigh wave speed of material~$n$ (for material~2, provided that $c_{R2}\leq c_{s1}$), despite the fact that $A_{II}^{(n)}(c\rightarrow c_{Rn})$ displays a simple pole. The reason for this, as it is easy to show, is that the divergences of $A_{II}^{(n)}$  are canceled by the factor $1-(a_n/b_n)W^2(c_{Rn})=0$  at  $c=c_{Rn}$. This ensures the regularity of $G_n$ when the rupture velocity crosses the Rayleigh wave speed of each material.

\item For bimaterial contrasts for which $c_{GR}$ exists, one has $W^2(c_{GR})\rightarrow +\infty$; since $c_{R1}\leq c_{GR}\leq \mathrm{min}(c_{R2},c_{s1})$, one has $G_1\rightarrow+\infty$ and $G_2\rightarrow-\infty$ for $c\rightarrow c_{GR}$. One also finds that $G_{\mathrm{sep}}\rightarrow \pm\infty$ as $(c-c_{GR})\rightarrow 0^\mp$ (see Fig.~\ref{fig:Gs}). Hence, the bimaterial case behaves near $c_{GR}$ in the same way as the homogeneous case as  $c \rightarrow c_R$.

\item For bimaterial contrasts where the generalized Rayleigh wave speed does not exist, no divergence of the energy release rates occurs. In addition, one can show that at $c=c_{s1}$ the inequalities $G_1>0$, $G_2<0$ and $G_{\mathrm{sep}}>0$ are satisfied for any bimaterial mismatch satisfying the nonexistence of $c_{GR}$.

\end{itemize}

Summarizing the above, we find that $G_1>0$ for $c<c_{GR}$ (resp. $c<c_{s1}$) when $c_{GR}$ exists (resp. does not exist). Moreover, provided that $c_{GR}$ exists, the behavior of $G_{\mathrm{sep}}$ for a bimaterial interface near $c_{GR}$ is similar to that of a homogeneous interface near the Rayleigh wave speed. This feature, however, disappears when $c_{GR}$ is not defined and $G_{\mathrm{sep}}>0$ for all $c<c_{s1}$. The most important result, however, concerns the behavior of $G_2$ with the rupture speed. For {\it any} bimaterial mismatch, a simple root of the equation $G_2=0$ {\it always exists}. This root defines a critical speed $c=c_T$ given by the solution of the equation
\beq
a_2(c_T)W^2(c_T)-b_2(c_T)=0\;.
\label{eq:defcT}
\eeq
Thus, $G_2>0$ for $c<c_T$ and $G_2<0$ in the interval $[c_T,c_{s1}]$. In contrast to $c_{GR}$, $c_T$ is always defined (a solution of Eq.~(\ref{eq:defcT}) always exists): $c_{R1}<c_T<c_{GR}$ if $c_{GR}$ exists and $c_{R1}<c_T<c_{s1}$ otherwise. Finally, $c_T$ coincides with material's Rayleigh wave speed, when the interface separates two similar materials.

One can wonder if this critical behavior of $G_2$, the energy release rate associated to the \textit{stiff} material, with the rupture front speed corresponds to any interesting experimental effects. In Section~\ref{sec:criterion}, we propose physical arguments in favor of this statement.

\subsection{Critical speed of subsonic bimaterial frictional rupture}
\label{sec:criterion}

Let us first recall the state of the art related to the dynamics of frictional rupture along an interface separating two similar materials. Because of the system's up-down symmetry, this case can be exactly mapped to the configuration of Mode~II crack propagation in a homogeneous material~\cite{Barras2019} (see for instance Eq.~(\ref{eq:Ghom})). Therefore, both fracture and homogeneous frictional rupture share the same feature: in the subsonic propagation regime, the rupture front speed is bounded by the Rayleigh wave speed. The theoretical ground of this limiting velocity is twofold: Stroh formalism~\cite{stroh1957theory} and energy considerations~\cite{freund1998dynamic,adda1999dynamic,Adda1999}. The former considers a rupture front as a propagating perturbation whose dispersion relation corresponds to waves propagating on the free surface of a semi-infinite elastic medium~\cite{rayleigh1885waves}. The latter is related to the \textit{positiveness} of the energy release rate, as the crack tip motion balances the energy provided by the elastic media with \textit{dissipative} mechanisms associated with the fracture energy. Perhaps coincidentally, both approaches predict the Rayleigh wave speed as the upper bound for crack tip dynamics.

For a bimaterial interface, the equivalent Stroh formalism suggests that the limiting speed is the generalized Rayleigh wave speed, provided that it is defined~\cite{stoneley1924elastic,weertman1963dislocations,gol1967surface}. This result is consistent with {\it naive} energy considerations for bimaterial frictional rupture dynamics; it satisfies the condition $G_{\mathrm{sep}}>0$, yielding  a limiting speed given by $c_{GR}$ when it exists, and $c_{s1}$ otherwise (see Fig.~\ref{fig:Gs}). This insight was suggested in~\cite{willis1971fracture}, in the context of interfacial crack propagation. 

We, however, now argue that $G_{\mathrm{sep}}$ \textit{is not} the only relevant energy release rate which governs rupture dynamics. Since Eq.~(\ref{eq:finalGn}) shows that $G_1\neq G_2$ for a bimaterial interface, one must now account for the potential impact of their behavior \textit{separately}. Recall that the interface separating two bulk materials is a surface of discontinuity of both elastic fields and material properties. The mechanical energy flowing through the upper and lower discontinuity surface can be transferred from one material to another along the entire interface $y=0$, except in the vicinity of the rupture tip at $x=\ell(t)$. The rupture tip is a physical \textit{singularity} where either material separation occurs, or contacts composing the interface are broken, to enable the bodies' relative motion. This irreversible breaking process inhibits any transfer of energy flows between upper and lower materials near the rupture tip. Consequently, the contributions to the mechanical energy flowing to the tip of both the soft and stiff materials should behave as distinct \textit{energy release rates}; they both must inject energy to generate propagation of the rupture front. Therefore, {\it both} $G_1$ and $G_2$ must, separately, satisfy the positiveness conditions
\beq
G_1>0 \quad\mbox{and} \quad G_2>0\;.
\label{eq:criterion}
\eeq

Equation~(\ref{eq:criterion}) is more restrictive than the global condition $G_{\mathrm{sep}}=G_1+G_2>0$. For a homogeneous system, Eq.~(\ref{eq:Ghom}) automatically holds; the two conditions are equivalent because of system's up-down symmetry. For the bimaterial case, the condition~(\ref{eq:criterion}) is fulfilled up to the critical rupture speed $c_T$ given by Eq.~(\ref{eq:defcT}). $c_T$ is defined for any bimaterial contrast, unlike the generalized Rayleigh wave speed. From this perspective, $c_T$ has the same physical origin as the Rayleigh wave speed for frictional rupture within similar materials. There are, however, two fundamental differences. First, the forbidden interval, $c_T<c<c_{s2}$ for bimaterial rupture stems from the behavior of the energy release rate of the stiff material only. Second, $G_2\sim (c-c_T) K^2$ near the critical speed, contrary to the homogeneous case where $G_{\mathrm{sep}}^{(\mathrm{hom)}}\sim K^2/(c-c_R^{(\mathrm{hom)}})$ near the Rayleigh wave speed. The behavior of $G_2$ near $c_T$ bodes well for a weak instability mechanism and suggests that either the propagation of frictional ruptures with a velocity $c\in[c_T,c_{s1}]$ is physically impossible, or that the elastodynamic problem as defined by Eqs.~(\ref{eq:bulk}--\ref{eq:bc-tau}) becomes ill-posed for $c>c_T$. In other words, the critical speed $c_T$ could be either a limiting velocity or a threshold indicating a transition to a different dissipative process at the ruptured interface. In the light of the following numerical and experimental results, we conjecture that both scenarios are possible and their occurrence depends on the propagation direction.

It is noteworthy that, up until now, the dynamics of the rupture front have been invoked only through the instantaneous speed $c(t)$. Indeed, the asymptotic stress field and energy release rates $G_n$ depend on the detailed tip dynamics only through the instantaneous stress intensity factor $K(t)$. This is not the case of $G_{\mathrm{slip}}$ which explicitly depends on the history of rupture propagation through the slip velocity $\partial\delta/\partial t$. 

The constitutive equations for the material response do not include the possibility of material separation. Therefore, the common way to include the separation process is to supplement the elastodynamic study with growth criteria. For both bimaterial and homogeneous frictional rupture, the usual criterion is to separate the dissipation process into two different contributions: $\Gamma_{\mathrm{sep}}$ and $\Gamma_{\mathrm{slip}}$. The fracture energy $\Gamma_{\mathrm{sep}}$ represents the resistance of the material to the rupture tip advance and must balance $G_\mathrm{sep}$, the usual Griffith criterion~\cite{freund1998dynamic}. On the other hand, the energy dissipated through sliding $\Gamma_{\mathrm{slip}}$ is an output of the problem computed from Eq.~(\ref{eq:Gslide}), provided that the residual stress $\tau_R$ is specified a priori. This approach is evidently valid only if one assumes a separation of scales between the dissipation involved in `breaking' contacts and in sliding, which might not always be the case~\cite{Barras2019}. Testing the validity of this approach is out of the reach of the present purely elastodynamic study, however the results we have presented so far are independent of the precise equation of motion of the rupture front. In the following, we will use both numerical approaches and experiments to reinforce our interpretation of the theoretical findings.

\section{Comparison with numerical results}
\label{sec:numerics}

We now develop a numerical model for the propagation of frictional rupture along bimaterial interfaces. The model is based on the spectral-boundary-integral method~\cite{geubelle1995spectral,breitenfeld1998numerical}, which solves the elastodynamic equations for a half-space. An explicit time integration is applied. The spectral description of the elastodynamic equations leads to a periodic set-up. We couple two half-spaces of different material properties by a cohesive-type approach along the frictional interface. Initially, the system is loaded by a uniform compressive normal load $N$ and static shear load $\tau_s$, and remains in equilibrium. We, then, artificially weaken the interface within an increasingly large area to create a `seed' crack. Once the seed crack is larger than a critical length, it becomes unstable and propagates dynamically along the interface. We then follow the rupture propagation and analyze its dynamic behavior. The nucleation process is both slow ($<0.1 c_{s1}$) and smooth enough to prevent any influence on the propagation.

The material of the two half spaces is linearly elastic and the properties correspond to the dynamic values of the materials used in the experiments described in Sec.~\ref{sec:experiments}. The upper material has elastic modulus $E_1 = 2.9~\mathrm{GPa}$, Poisson's ratio $\nu_1 = 0.39$, and density $\rho_1 = 1200~\mathrm{kg}/ \mathrm{m}^{3}$. The lower material has $E_2 = 5.65~\mathrm{GPa}$, $\nu_2 = 0.33$, and $\rho_2 = 1180~\mathrm{kg}/\mathrm{m}^{3}$. We apply plane-stress assumptions to approximate the thin blocks of the experimental set-up. 

Two different types of interface laws were considered. The first one, which we refer to as \textit{fracture}, is independent of the normal stress. The tractions within the rupture interface are governed by a linear slip-weakening cohesive law given by
\begin{equation}
\tau (\delta) = \begin{cases} (\tau_P - \tau_R) (1 - \delta / d_c) + \tau_R & \mathrm{for} \quad\delta < d_c \\ \tau_R &\mathrm{for} \quad \delta \geq d_c \end{cases}\;,
\label{eq:simfracture}
\end{equation}
where $\tau_P$ and $\tau_R$ are the peak and residual strength, respectively, $\delta$ is the tangential relative displacement also called local slip (see Eq.~(\ref{eq:bcdelta})), and $d_c$ is a characteristic length scale. This interface law corresponds to widely-used cohesive laws~\cite{camacho1996computational} with the slight difference of being applied to the tangential component of the interface tractions. Beyond the crack's tip, the tangential displacement is continuous across the interface, $\delta=0$, and the shear stress should satisfy $\tau < \tau_P$. 

The second type of interface law, which we call \textit{friction}, is similar to the fracture law, but with a dependence on the normal stress. The purpose of this approach is to test the implications of friction, which is macroscopically known to depend on the normal load. Therefore, using a Coulomb-like model, the local friction coefficient is given by
\begin{equation}
\mu (\delta) = \begin{cases} (\mu_s - \mu_k) (1 - \delta / d_c) + \mu_k &\mathrm{for} \quad \delta < d_c \\ \mu_k &\mathrm{for} \quad \delta \geq d_c \end{cases}\;,
\label{eq:simfriction}
\end{equation}
where $\mu_s$ and $\mu_k$ are the static and kinetic friction coefficients. The interface strength inside the crack is then computed by $\tau (\delta) = \mu (\delta) \sigma$, where $\sigma(x,t)$ is the normal stress, or contact pressure, at the interface. For stability reasons~\cite{cochard2000fault,andrews1997wrinkle,ben2002dynamic,rubin2007aftershock}, we do not apply directly this strength, but introduce a regularized version $\tilde \tau$ of it, which is governed by $\partial \tilde \tau / \partial t = (\tau - \tilde \tau ) / T^*$. However, we chose values for the time scale $T^*$ that are large enough to prevent stability issues but small enough to avoid perturbative effects on the quantities of interest for the present study~\cite{kammer2014existence}. Finally, we note that in everything that we show, there is displacement continuity in the normal direction. Hence, the normal pressure is continuous as well.

\begin{figure}[tb]
\begin{center}
\includegraphics[height=0.3\linewidth]{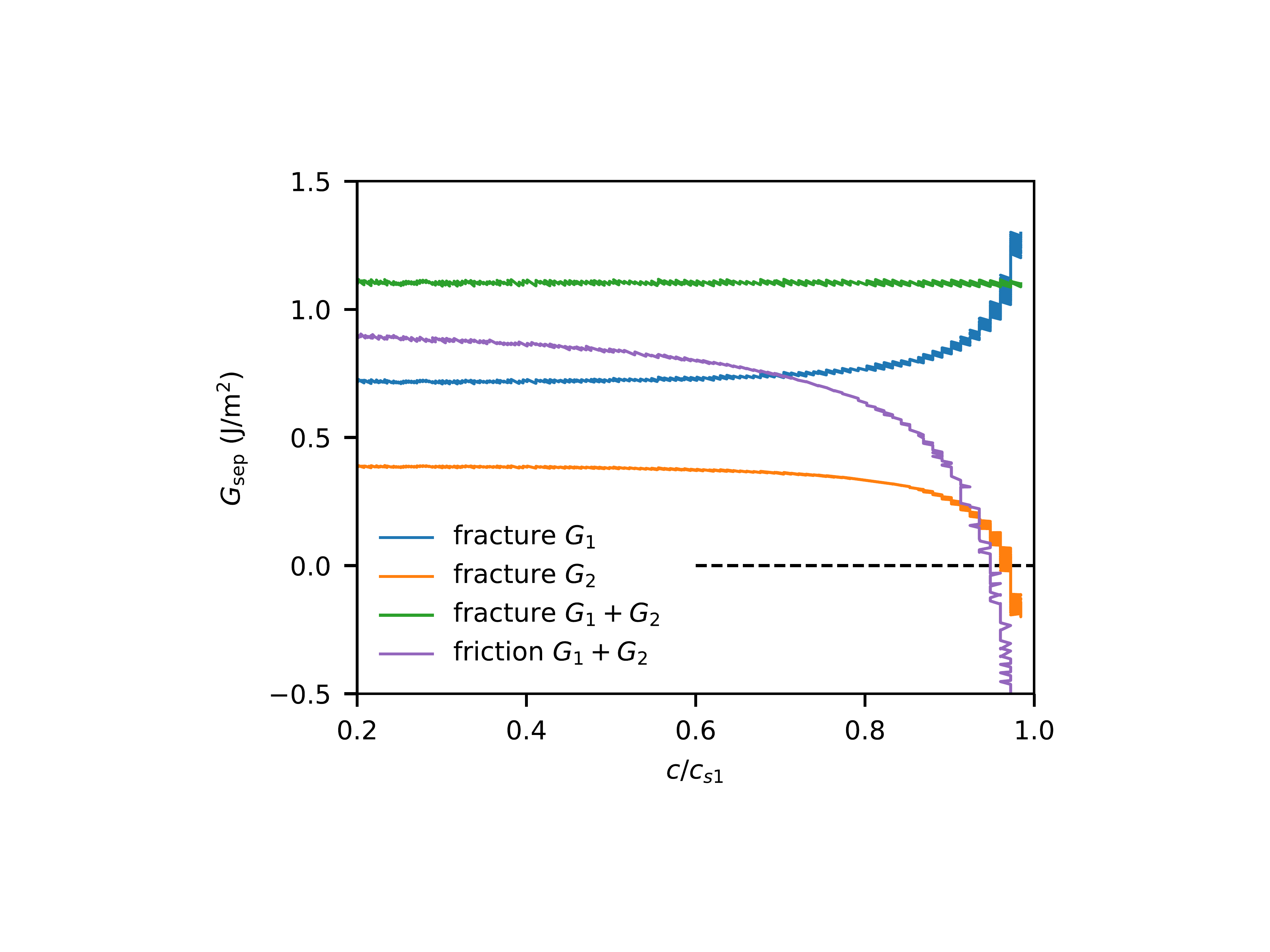}
\includegraphics[height=0.3\linewidth]{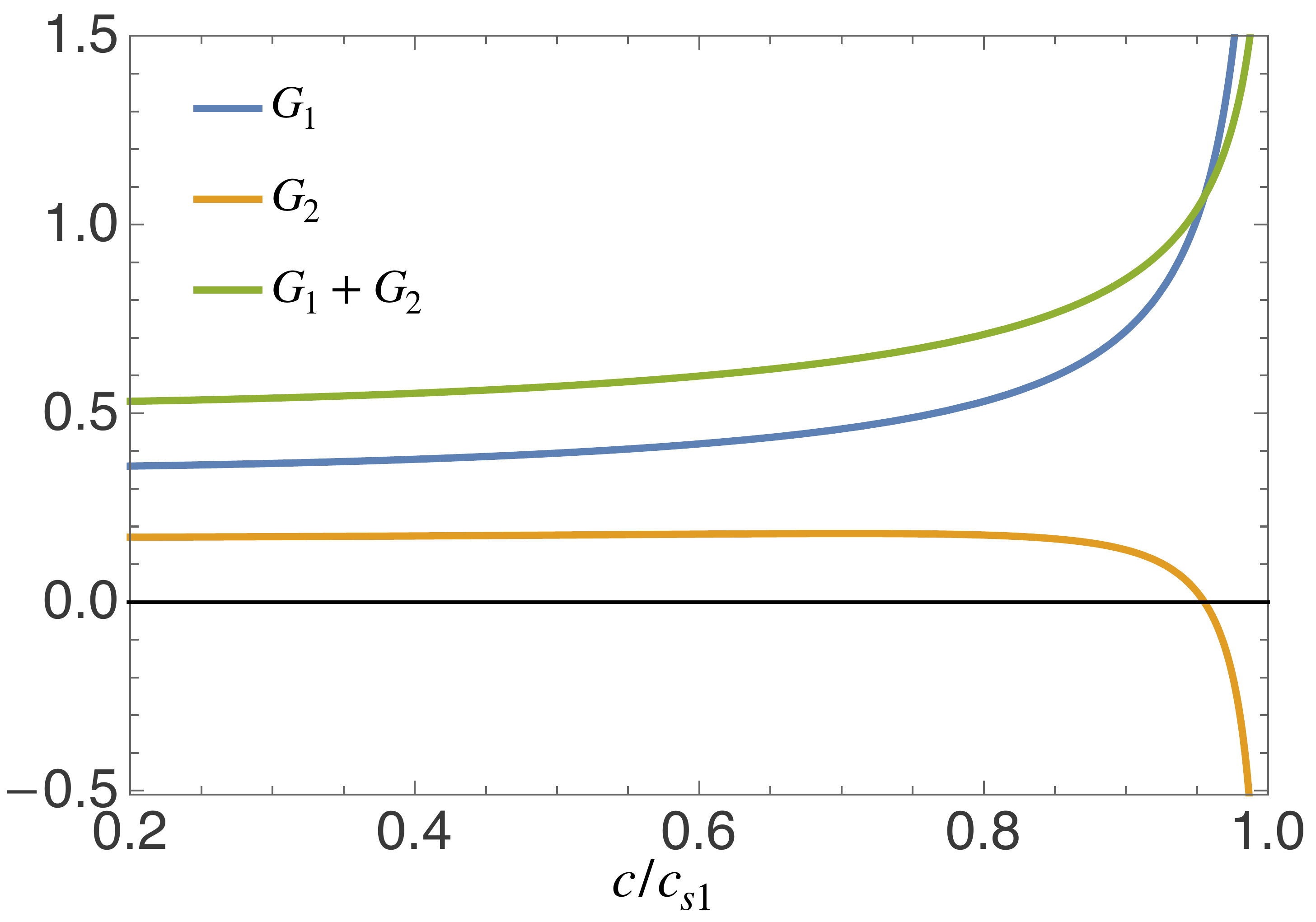}
\caption{Left: Energy release rates as function of the rupture speed computed numerically for both types of interfacial laws given by Eq.~(\ref{eq:simfracture}) for `fracture' and Eq.~(\ref{eq:simfriction}) for `friction'. For fracture, both $G_n$ and $G_{\mathrm{sep}}$ are shown while for friction only $G_{\mathrm{sep}}$ is plotted. Notice that $G_{\mathrm{sep}}$ for friction and $G_2$ for fracture vanish at approximately the same rupture speed. Bulk bimaterial parameters are the ones used in experiments of Sec.~\ref{sec:experiments} where plane-stress conditions are assumed. Right: theoretical dimensionless energy release rates $G_n$ and $G_{\mathrm{sep}}$ (scaled by $K^2/2\mu_1$) for the same bimaterial elastic parameters recovers the numerical results for `fracture' interface law: a transition occurs at a rupture tip speed $c_T=0.955c_{s1}$ ($c_T=867~\mathrm{m\cdot s}^{-1}$) governed by the condition $G_2=0$. The case where the interface law is of type `friction' exhibits the same feature but for $G_{\mathrm{sep}}$. In all numerical simulations, the magnitude of the remote applied loadings were $N= 5~\mathrm{MPa}$ and $\tau_s = 3.4~\mathrm{MPa}$. We used parameters of the `fracture' interfacial law: $\tau_P = 4.8~\mathrm{MPa}$,  $\tau_R = 3.2~\mathrm{MPa}$ and  $d_c = 1.37~\mu\mathrm{m}$, and for the `friction' case: $\mu_s = 0.96$, $\mu_k = 0.64$, $d_c = 1.37~\mu\mathrm{m}$ and $T^* = 0.25 \mu\mathrm{s}$.}
\label{Figure1_sim}
\end{center}
\end{figure}

The purpose of our numerical simulations was to compute the energy release rates defined in Sec.~\ref{sec:theory} using the interface laws given by Eqs.~(\ref{eq:simfracture},\ref{eq:simfriction}) and correlate them with the behavior of traction distributions along the sliding interface. The focus was on subsonic rupture fronts that propagate in the positive direction only. Subsonic propagation in the negative direction was hard to achieve within our numerical scheme. Within the system under study, the energy release rates $G_n$ of the top and bottom interfaces and the energy release rate $G_{\mathrm{sep}}$ used for material separation are given by~\cite{palmer1973growth,rice1980mechanics,freund1998dynamic}
\begin{eqnarray} 
G_{1}&=&\frac{1}{c} \int_{{\mathcal L}_1} \left[(\tau(\delta)-\tau_{R})\frac{\partial u^{(1)}_x}{\partial t}(x,0^+,t)-(\sigma(x,t)-\sigma_R)\frac{\partial u^{(1)}_y}{\partial t}(x,0^+,t)\right]dx\;,\\
G_{2}&=&-\frac{1}{c} \int_{{\mathcal L}_2} \left[(\tau(\delta)-\tau_{R})\frac{\partial u^{(2)}_x}{\partial t}(x,0^-,t)-(\sigma(x,t)-\sigma_R)\frac{\partial u^{(2)}_y}{\partial t}(x,0^-,t)\right]dx\;,\\
G_{\mathrm{sep}}&=&G_1+G_2= \frac{1}{c} \int_{-\infty}^{\ell(t)} (\tau(\delta)-\tau_{R})\frac{\partial \delta}{\partial t}(x,t)\,dx\;.
\label{eq:Gsepnumrics}
\end{eqnarray}
Here the notations introduced in Sec.~\ref{sec:theory} are used. Note that due to the addition of a cohesive zone at the tip, the fields are nonsingular at the rupture front, so there is no contribution from the integration over the contours ${\mathcal C}_n$ as in Eq.~(\ref{eq:Gsep}). Notice that in the definition of $G_n$, the normal contact pressure should also be subtracted from local normal stress component $\sigma(x,t)$ as is done for the shear stress component. For fracture $\tau_R$ is the residual shear stress appearing in Eq.~(\ref{eq:simfracture}) and $\sigma_R=N$, where $N$ is the applied pressure at the remote boundaries. However, for friction $\tau_R=\mu_k\sigma_R$ and $\sigma_R=N$ is still given by the applied normal load.

Figure~\ref{Figure1_sim} shows the resulting energy release rates of the bimaterial system used for the experiments reported in Sec.~\ref{sec:experiments}. When the interface law is described by Eq.~(\ref{eq:simfracture}), both the numerics and theory  coincide, showing that the introduction of a process zone does not alter the energy budget of the dissipation near the rupture front. Specifically, one recovers the functional behavior of $G_n$ with the rupture velocity and finds that, in contrast to the theoretical $G_{\mathrm{sep}}$, the value computed numerically is independent of the rupture speed. The latter result is expected since, provided that $\partial \delta/\partial t\approx -c\partial \delta/\partial x$, Eq.~(\ref{eq:Gsepnumrics}) yields $G_{\mathrm{sep}}\approx (\tau_P-\tau_R) d_c/2$ while the theoretical dimensionless $G_{\mathrm{sep}}$ does not include the contribution of the dynamic stress intensity factor $K(t)$. Nevertheless, the numerical results using an interface law given by Eq.~(\ref{eq:simfracture}) do not show any change in the behavior of the dissipation above the transition speed $c_T$, as conjectured in the theoretical study. The situation changes when one considers an interface law given by Eq.~(\ref{eq:simfriction}). Figure~\ref{Figure1_sim} shows that the energy release rate provided for material separation (breaking contacts) $G_{\mathrm{sep}}$ becomes negative for rupture speeds $c\in [c_T,c_{s1}]$. This result is more dramatic than the `fracture' case where only $G_2$ exhibits such behavior. Recall that the decomposition of energy flow into two contributions $G_{\mathrm{sep}}$ and $G_{\mathrm{slip}}$ supposes a separation of scales between the dissipative processes. Material separation (breaking contacts) occurs in the vicinity of the rupture front and sliding dissipation is involved along the whole interface.  Therefore, the numerical results for the `friction' interface law show that such separation no longer holds for rupture speeds $c>c_T$.

\begin{figure}[tb]
\begin{center}
\includegraphics[width=0.45\linewidth]{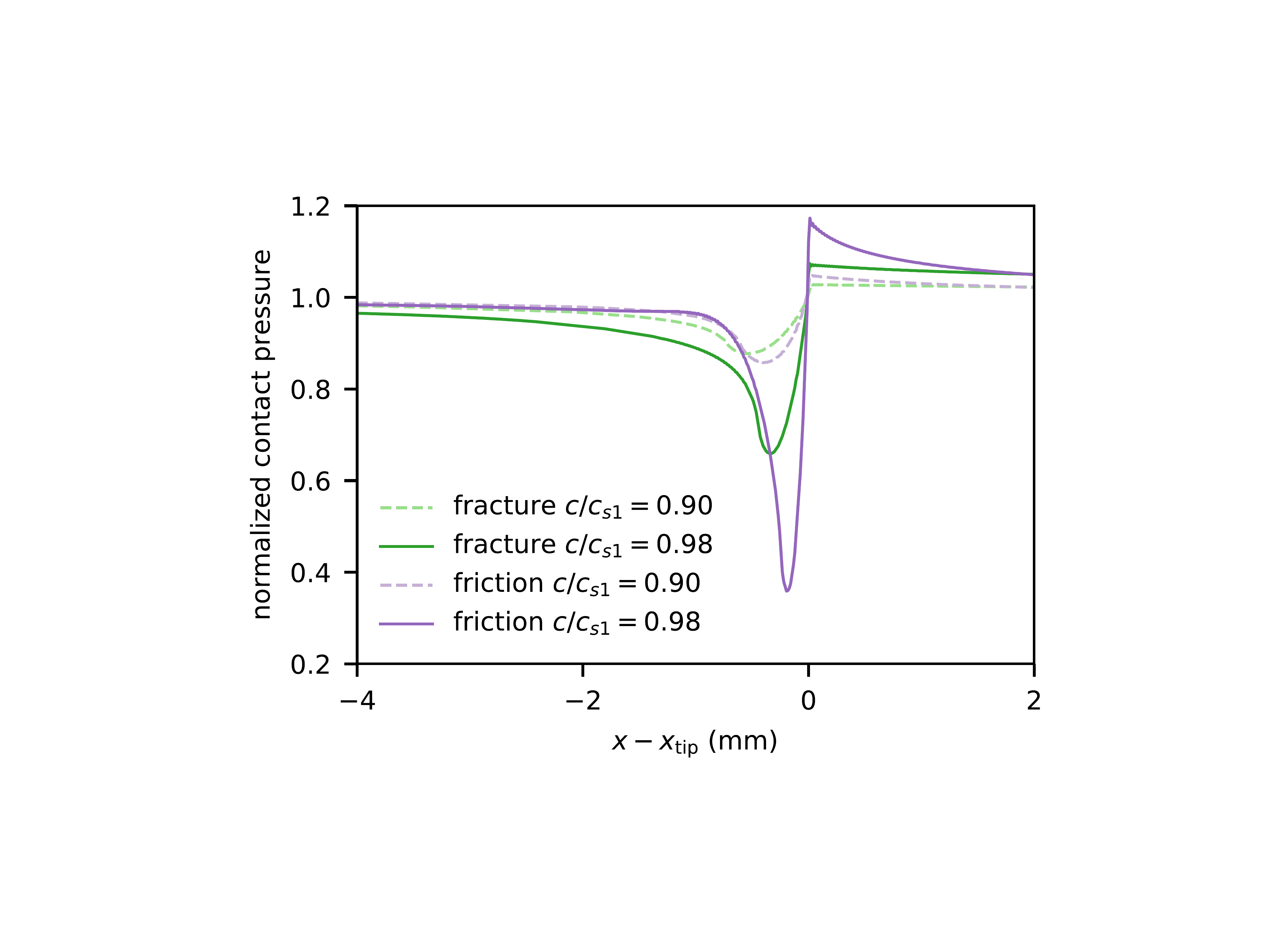}
\caption{Distribution of normal contact pressure along the interface for rupture speeds below and above $c_T=0.955c_{s1}$.  When $c<c_T$, both `fracture' and `friction' interface laws behave similarly. For $c>c_T$, the enhancement of bimaterial coupling effect is more pronounced for `friction' than `fracture' interface law. The large localized normal pressure drop for the `friction' interface law behind the rupture is a signature of slip-pulse nucleation when the rupture speed exceeds $c_T$.}
\label{Figure3_sim}
\end{center}
\end{figure}

To explore how loss of scale separation between the dissipative processes manifests itself along bimaterial interfaces, we study the behavior of the normal contact pressure when the rupture speed crosses $c_T$. The most important characteristic difference exhibited by bimaterial frictional rupture compared to homogeneous interface rupture is the coupling between the normal traction and slip rate along bimaterial interfaces. Figure~\ref{Figure3_sim} demonstrates that both `friction' and `fracture' interface laws behave similarly for rupture speeds where $c<c_T$: the interface contact behind the rupture front is only slightly reduced. However, when $c>c_T$ the drop of contact pressure is much more pronounced for the `friction' interface law compared to the `fracture' case. Including frictional coupling produces both a sharp localized reduction of the contact pressure behind the rupture front and an increase of contact pressure ahead of the front. Both of these features are characteristics of experimentally observed slip pulses~\cite{shlomai2016structure,shlomai_PNAS}. This observation suggests that, above $c_T$, a new mode of frictional rupture, which has the same characteristics as a slip-pulse, takes place. Therefore, we identify $c_T$ as the transition speed to a slip-pulse rupture mode in the positive direction.

These observations demonstrate the general validity and robust character of the general theoretical arguments developed in Sec.~\ref{sec:theory} with respect to the details of the interfacial law. These results are also in agreement with previous studies, where it was shown that bimaterial slip-pulses could result solely from bimaterial coupling; the form of the friction law is unimportant~\cite{weertman1980unstable,andrews1997wrinkle,ben2001dynamic,rice2001rate}. In the following, we turn to experiments for comparison to our theoretical and numerical findings.

\section{Comparison with experimental results}
\label{sec:experiments}

\subsection{Experimental system}

Here we provide a brief description of the system used in our measurements.  The same experimental set-up was used in ~\cite{shlomai_PNAS}, where a detailed description of our experimental apparatus and methods can be found. Our bimaterial interface was composed of a polycarbonate (PC) block of $(x,y,z)$ dimensions 197mm$\times$100mm$\times$5.8mm sliding on a 220mm$\times$100mm$\times$5.5mm PMMA block, with $x$ the sliding direction and $y$ the direction normal to the interface. The contacting faces of both blocks were diamond-machined to optical flatness. We measured the longitudinal ($c_{dn}$) and shear ($c_{sn}$) waves speeds under plane strain conditions and converted the measurements to the plane stress conditions that correspond to the dimensions of our sliding blocks. The corresponding (plane stress) wave speeds are $c_s^{PMMA} \equiv  c_{s2} = 1361\pm 13~\mathrm{m\cdot s}^{-1}$, $c_d^{PMMA} \equiv  c_{d2}  = 2345\pm 13~\mathrm{m\cdot s}^{-1}$, $c_s^{PC} \equiv   c_{s1} =908\pm 2~\mathrm{m\cdot s}^{-1}$ and $c_d^{PC} \equiv  c_{d1} = 1653\pm4~\mathrm{m\cdot s}^{-1}$. The mass densities, $\rho^{PMMA}\equiv\rho_2 =1170~\mathrm{kg\cdot m}^{-3}$ and $\rho^{PC}\equiv \rho_1=1200~\mathrm{kg\cdot m}^{-3}$ coupled to the wave speed measurements, yield dynamic values for the Poisson ratios of $\nu^{PMMA}\equiv\nu_2=0.33$ and  $\nu^{PC}\equiv\nu_1=0.39$. Both PMMA and PC are viscoelastic materials. Their (dynamic) Young’s moduli, relevant for the time scales of these experiments, were $E_{dynamic}^{PMMA} =5.75$GPa and $E_{dynamic}^{PC} =2.76$GPa.  

For each experiment, a fixed normal force of $2000$N$< F_N < 6000$N was imposed. An external shear load, $F_S$, was then quasistatically applied, as described in~\cite{shlomai_PNAS,shlomai2016structure}. Throughout each experiments the real contact area, $A(x,z,t)$, was monitored along the entire interface using an optical method based on total internal reflection~\cite{rubinstein2004detachment} where an incident sheet of light illuminated the frictional interface at an angle well beyond the critical angle for total internal reflection. The light was, therefore, transmitted through the interface only at the contacting points. The transmitted light was imaged at 580,000 frames per second with a spatial resolution of $x \times z$ = 1280$\times$8 pixels. These measurements  provided instantaneous values of $A(x,t) = \left\langle A(x,z,t) \right\rangle _z$ along the entire 1D interface.

We performed local strain tensor measurements, $\varepsilon_{ij}$, using miniature Kulite B/UGP-1000-060-R3 rosette strain gauges. 30 such gauges were mounted at heights, 3.5mm and 7mm, both above and beneath the interface. Each rosette strain gauge is composed of three independent active regions (each 0.4mm$\times$0.9mm in size) located in a 1mm$^2$ area to provide independent measurements of each component of the 2D strain tensor $\varepsilon_{ij}$. Each strain signal (60 channels) was individually amplified and simultaneously acquired to 14 bit accuracy by an ACQ132 digitizer (D-TACQ Solutions Ltd) at a 1MHz rate. This provided a sensitivity of ~3$\mu$Strain in $\varepsilon_{ij}(t)$ measurements (a 0.3\% uncertainty in $\varepsilon_{ij}(t)$). 

Despite this relatively high precision, the overall accuracy of our strain measurements is only 10-20\%, because of calibration variations between different strain gauge rosettes. This lack of absolute accuracy did not affect the majority of our measurements, since we are generally interested in relative strain variations that were acquired at given strain gauges. To compare different strain gauges, we neutralized any variations in calibration by normalizing the strain gauge outputs relative to their initial values. Strain variations, were  denoted by $\Delta {\varepsilon}_{ij}$ (the $\Delta$ signifies subtraction of the initial values of $\varepsilon_{ij}$). Stress variations, for example,  $\Delta{\sigma}_{yy}$, are defined in the same way. 

\subsection{Previous observations}

Previous work~\cite{shlomai_PNAS} has, experimentally, revealed that: 
\begin{itemize}[leftmargin=*]

\item In both the positive and negative propagation directions, ruptures initiate as subsonic (bimaterial) cracks; for slow ($c<0.8c_{s1}$) ruptures, slip is spatially extended and strains at the rupture tip possess the $r^{-1/2}$ singularity characteristic of crack solutions.

\item In the positive direction, steady-state subsonic cracks are rarely observed. In the bimaterial system used in these experiments, rupture fronts generally rapidly accelerate through the subsonic regime, until reaching a transonic regime and obtaining an observed limiting velocity of $c^{lim}=1.041 c_{s1}$.  In the transonic regime {\it only}  slip pulses are observed. This limiting velocity, for the bimaterials system used, has been predicted theoretically~\cite{shlomai_PNAS}. 

\item In the negative direction, subsonic ruptures can be observed. They are, however, less prevalent than supershear ruptures which are the preferred propagation mode~\cite{shlomai2016structure,shlomai_jgr} with speeds concentrated in a narrow band that has been explained theoretically~\cite{shlomai_jgr}.

\end{itemize}

\subsection{Experimental support for $c_T$}

Here we focus on analyzing the form of ruptures as they transition through $c_T$ and eventually morph into slip pulses beyond $c_{s1}$. We first wish to compare the structures of strain fields, $\varepsilon_{ij}$, surrounding  the tips of both the negatively propagating and positively propagating fronts for relatively slow rupture velocities. Before we examine experimental data, we note that previous experiments~\cite{Svetlizky2014,shlomai_PNAS} have demonstrated that for propagation speeds that do not approach the limiting velocities, the forms of $\varepsilon_{ij}(t)$, when plotted as a function of the distance from the crack tip, $x- x_{tip}$, are fairly insensitive to the instantaneous crack speed. This rather useful feature of the LEFM solutions for both homogeneous~\cite{freund1998dynamic} and bimaterial~\cite{shlomai_PNAS} cracks enables us to perform a detailed comparison of the measured forms $\varepsilon_{ij}(x-x_{tip})$ for `slow' bimaterial ruptures in each direction with constant velocity predictions for bimaterial cracks - even when the experimental ruptures are accelerating. Moreover, we note that the analytic solutions obtained under the `{\it fracture}' boundary conditions (Eq.~\ref{eq:simfracture}) are insensitive to the propagation direction. This contrasts with the `{\it friction}' boundary conditions (Eq.~\ref{eq:simfriction}), where the variations of the normal stress at the interface are coupled both to frictional resistance through the friction law and to the slip velocity through the bimaterial coupling effect (see Eqs.~(\ref{eq:syy},\ref{eq:coupling})).

\begin{figure}[htb]
\begin{center}
\includegraphics[width=0.5\linewidth]{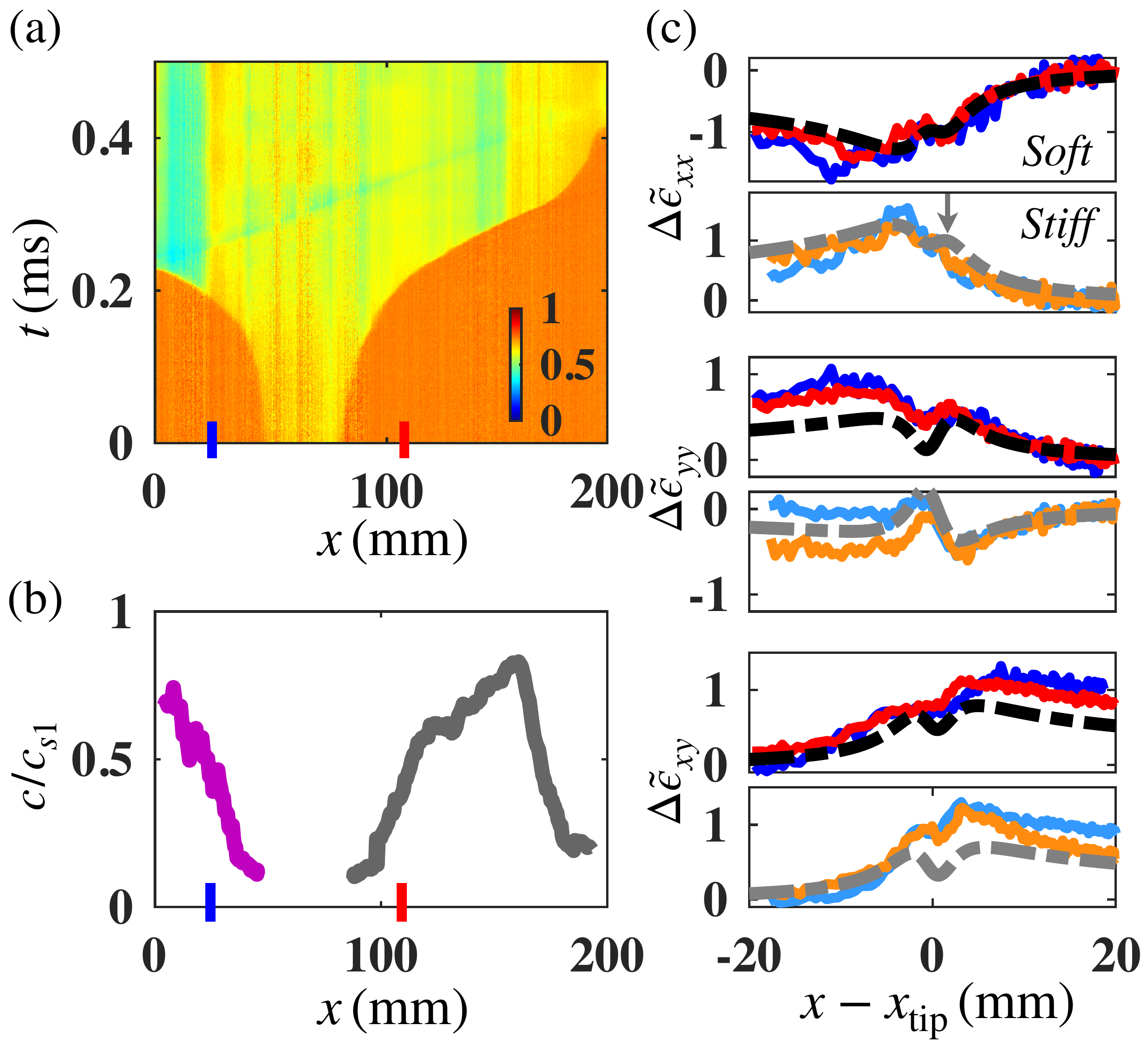}
\caption{Counter-propagating ruptures which were formed by a spontaneous nucleation at about $1/3$ of the interface length and developed to similar ruptures. a) $A(x,t)$ when normalized by $A(x,0)$, where $t = 0$ corresponds to an arbitrary point within the (slow) nucleation stage. b) Rupture velocity profiles as functions of the rupture locations along the interface. Grey (magenta) correspond to the rupture propagating in the positive (negative) direction. c) Normalized strains, $\Delta \tilde{\varepsilon}_{ij}$. Strains were normalized as follows;  strain variations, $\Delta \varepsilon_{ij}$, are obtained by subtracting either the initial ($yy$ and $xx$ components) or the residual ($xy$ component) values, then $\Delta \tilde{\varepsilon}_{ij} \equiv \Delta \varepsilon_{ij}/\Delta \varepsilon_{\mathrm{m}}$, where $\Delta \varepsilon_{\mathrm{m}}$ is the strain at the first peak of $\Delta \varepsilon_{xx}^{(2)}$. This peak occurs slightly ahead of the tip \cite{shlomai_PNAS} (see grey arrow). This normalization enables us to quantitatively compare the relative amplitudes of the different $\Delta \varepsilon_{ij}$.  The signal colors indicate both the block material and the measurement locations along the interface as noted in (a) and (b). The $x$ locations of the strain gauges on opposing blocks are located to within 3-4mm of each other. Darker colors (blue and red) belong to the softer material (PC) and lighter colors (light blue and orange) to the stiffer (PMMA) one. The signs of the $\Delta \tilde{\varepsilon}_{xx}$ and $\Delta \tilde{\varepsilon}_{yy}$ strain fields of ruptures propagating in the negative direction are opposite from those of ruptures propagating in the positive direction. To facilitate comparison, the signs of the strain fields propagating in the negative direction are reversed; $-\Delta \tilde{\varepsilon}_{xx}$ and $-\Delta \tilde{\varepsilon}_{yy}$. Measurements were performed at distance $h = \pm3$~mm from the interface. The maximal relevant distances from the rupture tip, $x-x_{tip}$, correspond to the physical distance along which the rupture could propagate before encountering the block boundaries. Dashed black and grey lines are the stiff and soft components, respectively, of the analytical solution at a constant velocity of $c=0.43c_{s1}$, corresponding to the local speed at the measurement points. Each of the stiff and soft components of the solution were normalized as the measurements. The deviation of the analytical solution from the measurements for $\Delta \tilde{\varepsilon}_{xx}$ and $\Delta \tilde{\varepsilon}_{xy}$ may be the result of the increasing magnitudes of the experimental strain fields with $c$, an effect not completely negated by normalization by $\Delta \varepsilon_{\mathrm{m}}$, as the ruptures are not in steady-state propagation.}
\label{Figure1_exp}
\end{center}
\end{figure}

In Fig.~\ref{Figure1_exp}(a), we present an example of fronts propagating in both directions within the same experiment. Here, rupture nucleation took place at an interior point ($x \sim 70$mm) along the interface and contact reduction of the interface was mediated by two counter-propagating fronts. The applied shear stress at the time of nucleation was, moreover, relatively low. As a result, in this rare case, both fronts propagated at relatively low speeds; accelerating from rest to about $\sim 0.7c_{s1}$ (see Fig.~\ref{Figure1_exp}(b)).

We are now in a position to directly compare the structures of strain fields, $\varepsilon_{ij}$, surrounding  the tips of both the negatively propagating and positively propagating fronts. This comparison is presented in Fig.~\ref{Figure1_exp}(c), where all components of the normalized $\Delta \tilde{\varepsilon}_{ij}$ measured at the tips of both fronts are shown. Measurements in both the stiff and soft materials are presented. The strain fields are compared both to each other and to the theoretical square root behavior. The results clearly show that all components of the strain fields, for ruptures in both directions, are nearly identical. Moreover, the measured $\Delta \tilde{\varepsilon}_{ij}$  are in very good agreement with the analytical predictions for slowly propagating `crack-like' rupture modes calculated for `{\it fracture}' boundary conditions.
 
\begin{figure}[htb]
\begin{center}
\includegraphics[width=0.5\linewidth]{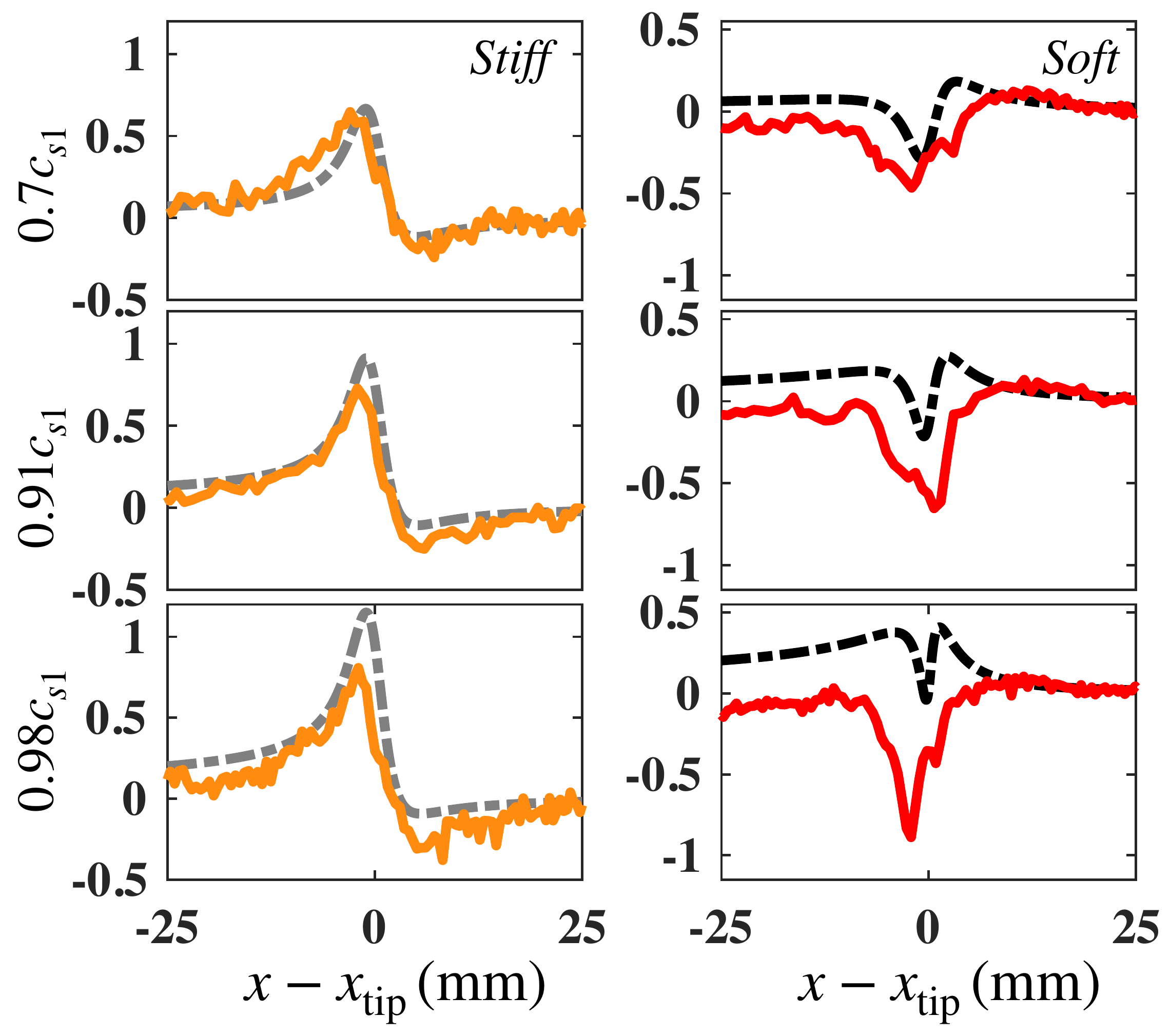}
\caption{{\it Positive} propagation direction: development of fast subsonic rupture structures with increasing propagation velocity $c$. Normalized experimental (solid curves) $\Delta\tilde{\sigma}_{yy}$ stress components are compared to bimaterial fracture-like solutions obtained by LEFM (dashed curves), for both stiff (left column within each subplot) and soft (right column within each subplot). Notice that the normalization of $\Delta\tilde{\sigma}_{ij}$ results from that of the strain field $\Delta \tilde{\varepsilon}_{ij}$ as explained in Fig.~\ref{Figure1_exp}. Various high subshear velocities $c/c_{s1}=$ 0.7, 0.91 and 0.98 are presented in each row. $\Delta\tilde{\sigma}_{yy}^{(2)}$ signals are in good agreement with the analytical calculations.  For $\Delta\tilde{\sigma}_{yy}^{(1)}$ the analytical prediction significantly deviates from the measurements at even the lower velocities. The solution completely fails to match the measurements at the higher velocities. Note that  $c_T\approx 0.955c_{s1}$.  
}
\label{Figure2_exp}
\end{center}
\end{figure}

Let us now turn to higher velocity frictional ruptures in the {\it positive} direction. In Fig.~\ref{Figure2_exp} we compare the normal stress variations,  $\Delta\tilde{\sigma}_{yy}$, to theoretical predictions (fracture boundary conditions) for propagation velocities that straddle the transition velocity, $c_T\approx 0.955c_{s1}$. On both sides of the interface, the $\Delta\tilde{\sigma}_{yy}$ signals are in reasonable agreement with theory for $c=0.71c_{s1}= 0.75c_T$. As $c$ approaches $c_T$, at $c=0.91c_{s1}= 0.96c_T$, the $\Delta\tilde{\sigma}_{yy}$ signals already deviate strongly from the theoretical predictions. Beyond $c_T$ (at $c=0.98c_{s1}= 1.03c_T$), the deviations from the theory have significantly increased on the soft side of the interface. It is interesting that, on the stiff side of the interface, while there are increasing discrepancies between predicted and measured values of $\Delta\tilde{\sigma}_{yy}$, their magnitude is much less than on the soft side of the interface. While we have presented data for $\Delta\tilde{\sigma}_{yy}$, the same qualitative behavior is evident for all other stress and strain components; significant digressions from theory on the soft side of the interface as we transcend $c_T$ and smaller (but systematically increasing) deviations between theory and experiments on the stiff side. 
 
\begin{figure}[htb]
\begin{center}
\includegraphics[width=0.7\linewidth]{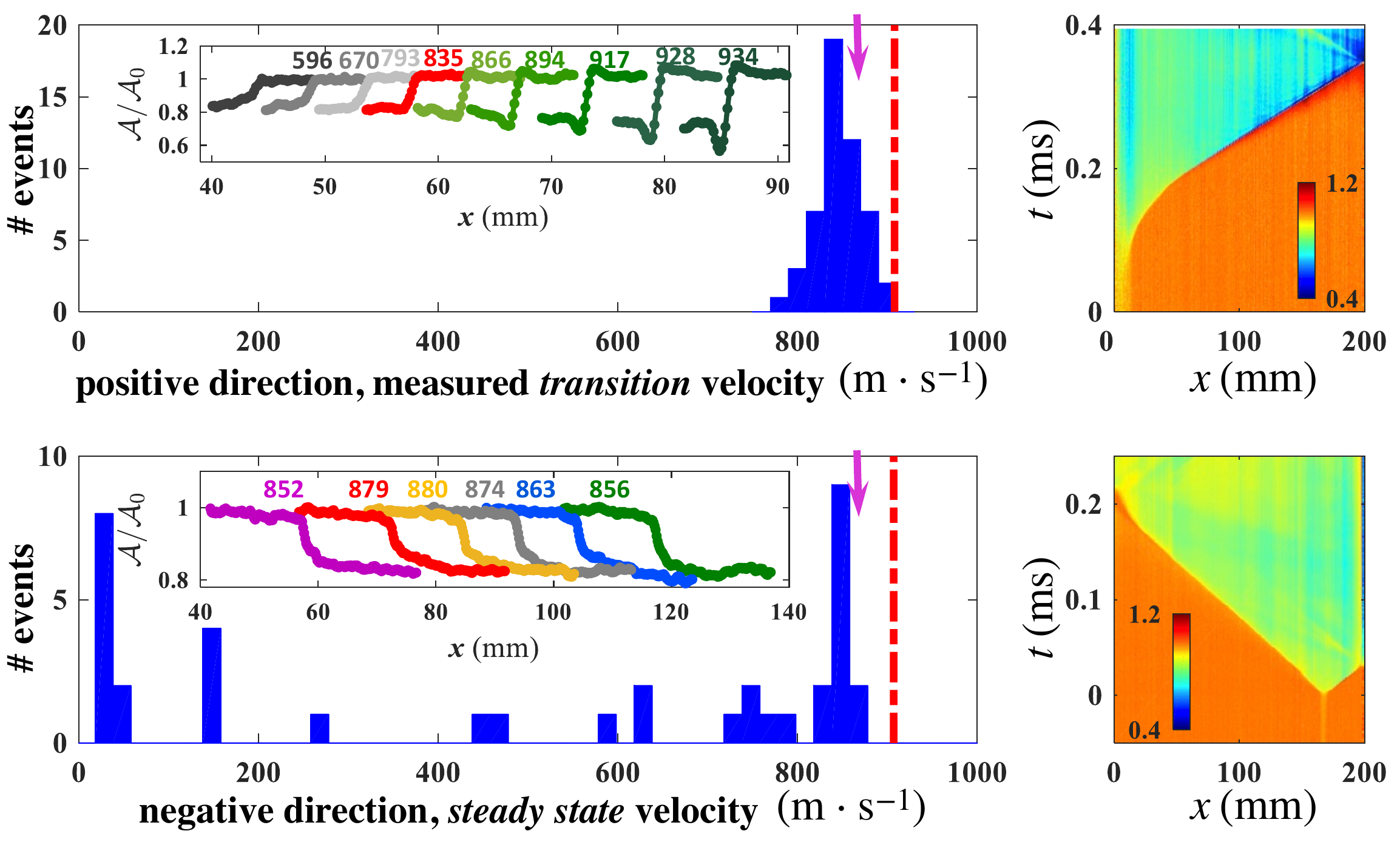}
\caption{(a) Measured transition velocities of ruptures propagating in the positive direction as the maximal rupture velocity for which $A(x,t)$ have the step-function like profile that is characteristic of crack-like ruptures. Beyond the transition velocity, contact area profiles (in the positive direction) exhibit increasingly non-monotonic features that are typical of slip pulses. (inset) a typical example of the development of $A(x,t)$ with rupture speed for the accelerating rupture shown in the right panel. Time increased from the left to the right. Propagation velocities for each profile are noted with the profile corresponding to the experimentally detected transition velocity in red. (b) Distribution of measured steady state rupture velocities in the negative direction. (inset) typical examples of step-like $A(x,t)$ profiles at different times in the experiment described in the right panel, time increased from the right to the left. Note that in the negative direction, no transition occurs and the rupture speed does not exceed the predicted value of $c_T$. In both (a) and (b) the red dashed line denotes the shear wave speed of the soft material $c_{s1}=908~\mathrm{m\cdot s}^{-1}$ and the predicted value of $c_T=867~\mathrm{m\cdot s}^{-1}$ is denoted by an arrow. Note that $x$-axes in both insets correspond to those in the right panels.
}
\label{Figure3_exp}
\end{center}
\end{figure}

While Fig.~\ref{Figure2_exp} is suggestive that a transition at $c_T$ indeed takes place, unfortunately we are technically rather limited in the continuum of propagation velocities for which we can compare the predicted to measured stresses.  The reason for this is that our strain gauge array locations are discrete, so comparisons similar to  Fig.~\ref{Figure2_exp} are only possible if a rupture with a desired velocity traverses one of these locations. We can use the contact area measurements, however, to circumvent this problem. $A(x, t)$ are, in a sense, a proxy for $\Delta\tilde{\sigma}_{yy}$ signals, as (to first order) they are approximately proportional to one another~\cite{bowden2001friction,Rubinstein2006}.  Moreover, $A(x,t)$ are measured (by definition) {\it on} the interface, whereas the normal stress measurements are displaced normal to the interface by a few mm's. In Fig.~\ref{Figure3_exp} we compare $A(x,t)$ profiles for increasing rupture velocities. We do this for both positively and negatively propagating ruptures. 

First, let us consider propagating ruptures in the positive direction. For $c < c_T$, both measurements of $A(x,t)$ and predictions for $\sigma_{yy}$ show that the normal pressure on the interface has an approximate step function form near the rupture front. Once $c > c_{T}$, however, ruptures start to approximate a slip pulse. Characteristic features of a slip pulse~\cite{shlomai2016structure,shlomai_PNAS} include a clear increase of the normal stress ahead of the rupture tip together with a pronounced and highly localized reduction of both $A(x,t)$ and the normal pressure behind the tip. These characteristic features are also observed in the simulations (Fig.~\ref{Figure3_sim}); both `fracture' and `friction' predictions have this `step function' form for $c<c_T$, whereas {\it above} $c_T$ the `friction' solution, in particular, starts to approximate all aspects of a slip pulse.  We can, therefore, use these characteristic features as a way to identify the transition from a crack-like rupture to `something else', for example a rupture mode developing into a slip pulse in the positive direction. In each of our numerous experiments, we define a `transition velocity' as  the velocity for which both an increase of $A(x,t)$ ahead of the rupture tip and a dip in $A(x,t)$ behind it first occur. In Fig.~\ref{Figure3_exp}(a) we present a histogram of the transition velocities of all of the measured events in the positive direction. The histogram is sharply peaked with a mean value, $c\approx 858~\mathrm{m\cdot s}^{-1}$, nearly exactly the theoretically predicted value of $c_T=867~\mathrm{m\cdot s}^{-1}$.

Let us now consider propagating ruptures in the negative direction. Here, there is no transition to a slip pulse, as slip pulses do not exist in the negative direction. We can still, however, perform the same type of analysis, now asking what is the highest subsonic propagation velocity in the negative direction for which the $A(x,t)$ possess the approximate step-function form that characterizes a crack-like rupture. These data are presented in Fig.~\ref{Figure3_exp}(b). Again, we observe a different type of `transition' at a velocity that is nearly identical to $c_T$. In this case, we find that, beyond $c_T$, {\it no} ruptures exist at all. There is a pronounced gap between the highest observed rupture velocity in the negative direction - occurring at the theoretically predicted value of $c_T$ and the shear velocity, $c_{s1}$.

As subsonic negatively propagating ruptures in bimaterial systems are not common (as can be seen by the dearth of events in Fig.~\ref{Figure3_exp}(b)), one could claim that there are insufficient statistics to determine $c_T$ to high precision from these data. We would argue, however, that at the very least, these data  highly support the proposition that $c_T$ indeed exists. Moreover, when coupled with the data from the positive direction, this analysis provides strong evidence that bimaterial cracks indeed lose their stability at the well defined value of $c=c_T$. Beyond this transition velocity, bimaterial cracks either transition to slip pulses (in the positive direction)  or simply disappear (in the negative one). The instability is caused by a single physical mechanism (negativity of $G_2$). Of course, what the unstable crack transitions to, is entirely different and, we believe, is governed by the sign of the bimaterial coupling at the unstable crack's tip. 

\section{Discussion}
\label{sec:discuss}

The numerical and experimental results unambiguously confirm that $c_T$ is, indeed, a threshold speed for bimaterial frictional rupture in both propagation directions.  We note that our theoretical prediction is based on a purely elastodynamic observation; the energy flowing from the stiff half-space becomes negative at $c_T$. This situation is, obviously an un-physical phenomenon; suggesting that the dissipative region must somehow act as a conduit to `feed' energy into the stiff one to maintain the singular field at the crack tip. While this calculation tells us that a crack-like solution must become unstable, it cannot tell us what type of rupture mode replaces this unstable mode of propagation. 

The instability of this solution, additionally, tells us something much more. The singular propagation mode for $c<c_T$ is the {\it only} dynamically propagating solution of the elastic wave equations subject to the boundary conditions assumed in Eqs.~(\ref{eq:bulk})-(\ref{eq:bc-tau}); contact is preserved at the interface for all $x$ and $t$ and that dissipative processes are separable as   $\Gamma_{\mathrm{sep}}$ and $\Gamma_{\mathrm{slip}}$ suggest.

In the positive propagation direction, both experimental and numerical observations demonstrate that a different {\it type} of rupture mode indeed exists along the interface beyond $c_T$. In~\cite{shlomai_PNAS} we observed that slip pulses are excited at velocities greater than $c_{s1}$. The present study suggests that slip pulse formation may actually initiate for $c>c_T$ (see e.g. Fig.~\ref{Figure3_exp}(a)). The fact that, beyond $c_T$, we `lose' the singular solution that exists for $c<c_T$ suggests that either one or all of the assumptions that we made are no longer valid. Our results suggest that the likeliest candidate is that the assumption of scale separation breaks down. The prediction of negative $G_2$ for cracks assumes a separation of scales in which all of the dissipative processes related to contact separation can be  encompassed within a small vicinity of a crack's tip (a `cohesive zone'). It  is  possible that this is exactly what is breaking down when slip pulses start to evolve.  As can be seen in Fig.~\ref{Figure3_sim}, `friction' cracks, in the vicinity of $c_T$, start to develop the reduced $\sigma_{yy}$ that is the hallmark of slip pulses.  If the size of the region of reduced $\sigma_{yy}$ becomes too large, then the scale separation of the dissipative regions may break down, as the regions of $G_{sep}$ and $G_{slip}$ may start to overlap.  While $G_{sep}$ may become negative above $c_T$, the {\it total} energy release rate, $G_{sep} +G_{slip}$ (for friction) will still remain positive. If  $G_{sep}>>G_{slip}$ then the boundary conditions in Eqs.~(\ref{ubc}-\ref{eq:bc-tau}) are valid. On the other hand,  if there is no scale where  $G_{sep}$  dominates $G_{slip}$, then  $G_{sep}<0$ cannot, in itself, be considered to be a stability condition.   
 
Such a process is not encoded in the LEFM framework, although both numerical and experimental results suggest that a large reduction of contact pressure behind the rupture front indeed signals the nucleation of a slip-pulse rupture mode. However, for such a process to occur, the bimaterial interface below the transition speed should be in a state that allows a reduction of contact pressure. This is precisely the case for propagation in the positive direction, where Eq.~(\ref{eq:syy}) shows that $K(t) W(c)>0$ for $c(t)\approx c_T$; the interface behind the rupture front, $x<\ell(t)$, is in a state that allows for `opening' because the contact pressure is lowered. Therefore, a slip-pulse mode of propagation can be nucleated at the critical rupture speed $c(t)=c_T$ and will continue to build up at speeds $c(t)>c_T$. Consequently, in the \textit{positive} direction $c_T$ is a \textit{transition} speed from fracture-like to pulse-like rupture modes. In addition, we would expect a slip-pulse to rapidly accelerate so long as $K(t) W(c)>0$, due to the positive feedback between increased rupture velocity and decreased contact pressure. The  condition $K(t) W(c)>0$ (postive feedback mechanism) is satisfied for rupture speeds up to $c_{GR}$, when it exists, and beyond $c_{s1}$ otherwise. As a result, when the generalized Rayleigh wave speed exists, one expects slip-pulses to accelerate up to an asymptotic speed $c^{lim}\equiv c_{GR}$ whereas, if $c_{GR}$ does not exist, the slip-pulse propagation regime could exceed $c_{s1}$. For the latter case, it was shown that slip-pulse propagation is, indeed, limited by a well-defined asymptotic transonic speed $c^{lim}>c_{s1}$~\cite{shlomai_PNAS} that corresponds to the point where the normal stress reduction behind the rupture front ceases to be positive.

In the negative propagation direction, the situation is apparently much simpler. In this case, experiments suggest that there are {\it no} other solutions beyond $c_T$, except for supershear modes~\cite{shlomai_jgr, shlomai2016structure}. The only scenario that can take place  in the negative direction is, therefore, that subshear cracks are either limited by $c_T$ or, if energetically possible, supershear solutions are nucleated. The mechanism for this behavior is seen as follows. Assume a dynamical bimaterial frictional rupture front accelerating in the \textit{negative} direction at a time-dependent speed $c(t)$. For rupture speeds up to $c(t)=c_T$, the bimaterial response is similar to that of the positive direction and is well described by the elastodynamic framework. However, the situation changes for possible rupture speeds $c(t)>c_T$.  Equation~(\ref{eq:syy}) shows that $K(t) W(c)<0$ for $c(t)\approx c_T$; the interface behind the rupture front, $x<\ell(t)$, is then in a state that inhibits slip-pulse nucleation because the contact pressure (hence the effective fracture energy) is enhanced. Therefore, the impossibility of changing to a qualitatively different type of rupture mode prevents frictional ruptures from exceeding the threshold rupture speed $c_T$. These arguments demonstrate why slip pulses are never observed in the negative direction~\cite{shlomai2016structure,shlomai_jgr}, the only other possible bimaterial rupture modes are supershear~\cite{shlomai_jgr}. Consequently, in the \textit{negative} direction $c_T$ is a \textit{limiting} speed for subsonic bimaterial frictional rupture propagation.

\begin{figure}[tb]
\begin{center}
\includegraphics[width=0.45\linewidth]{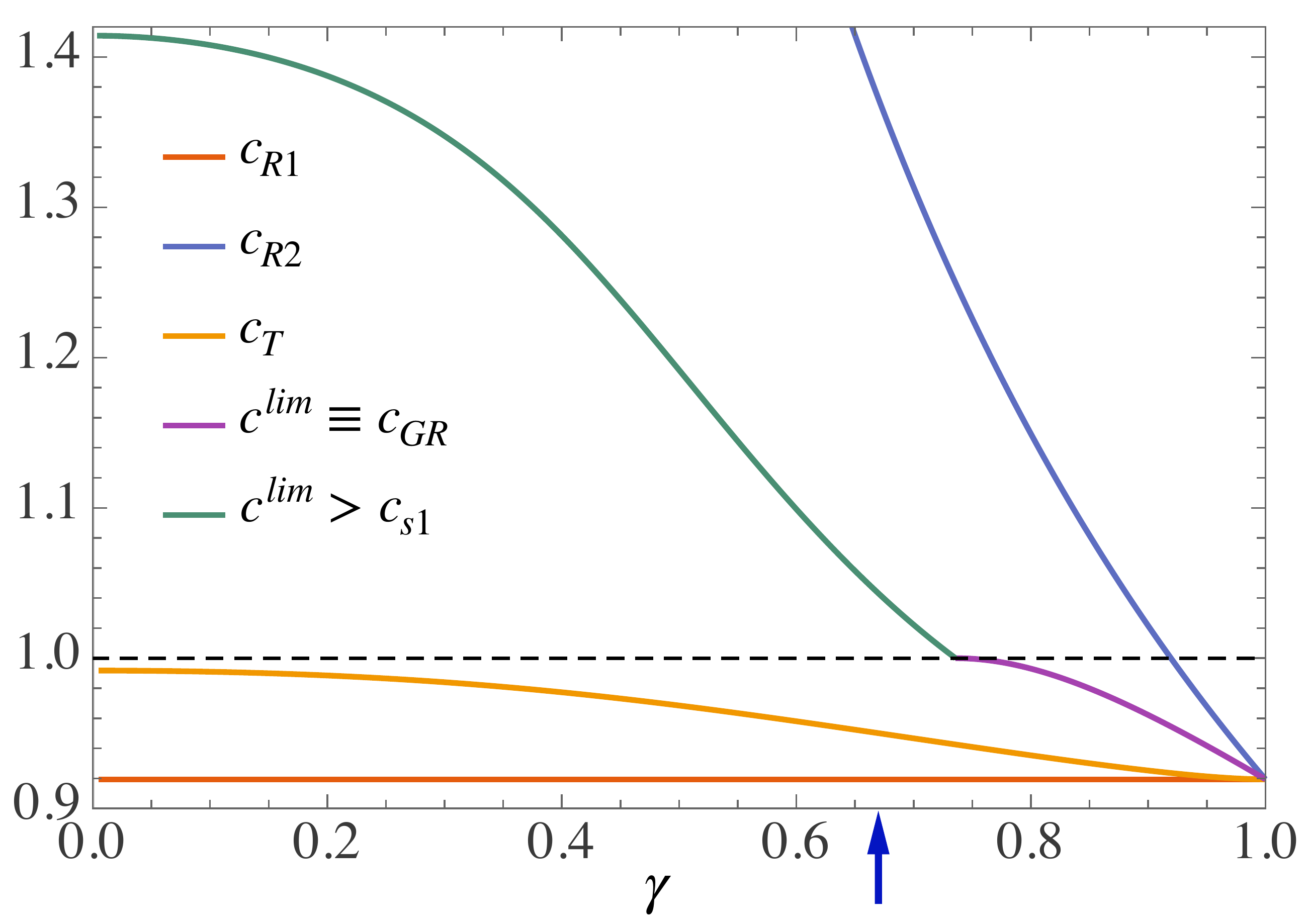}
\caption{The dimensionless critical speeds $c_T$ and $c^{lim}$ as function of bimaterial mismatch $\gamma=c_{s1}/c_{s2}$ for model bimaterials with densities $\rho_1=\rho_2$ and $c_{d1}/c_{s1}=c_{d2}/c_{s2}=\sqrt{3}$. Also shown are the characteristic wave speeds $c_{R1}$, $c_{R2}$ and $c_{GR}$, when it exists. All velocities are non-dimensionalized by $c_{s1}$. Recall that depending on the propagation direction, $c_T$ is both a \textit{transition} speed from fracture-like to pulse-like rupture modes (positive direction) and  a \textit{limiting} speed for subsonic bimaterial frictional rupture propagation (negative direction). On the other hand, $c^{lim}$ is the \textit{limiting} speed for slip-pulse rupture modes that only exist in the positive direction.  Note that, while $c_T$ is subsonic and exists for all bimaterials, one has $c^{lim}\equiv c_{GR}$ (i.e. subsonic) for moderate bimaterial contrast where $\gamma\geq0.736$, and $c^{lim}>c_{s1}$ (i.e. transonic) otherwise. For the case of a soft-infinitely stiff bimaterial ($\gamma=0$), it can be shown that $c^{lim}=\sqrt{2}c_{s1}$ while $c_T\leq c_{s1}$ depends on the Poisson ratio of the soft material. Finally, the blue arrow denotes the value of $\gamma\approx  0.67$, which corresponds approximately to the bimaterial contrast used in the experiments.}
\label{fig:Cs}
\end{center}
\end{figure}

Concluding our discussion, we describe in Fig.~\ref{fig:Cs} the general behavior of the transition speed $c_T$ and the asymptotic speed $c^{lim}$ for model bimaterials described by a single mismatch parameter $\gamma=c_{s1}/c_{s2}$ (in our experiments $\gamma \approx0.67$). This phase diagram demonstrates how the transition and limiting velocities are related as $\gamma$ varies. In particular it is seen that the limiting velocity continuously varies from $c_{GR}$ to transonic speeds, the two behaviors merging precisely at the point when  $c_{GR}$ ceases to exist.  The phase diagram further predicts that, for any $\gamma$, a band of slip pulses is possible for (positive) propagation velocities in the region bounded by $c_T$ and $c^{lim}$. 

\section{Conclusion}
\label{sec:conclusion}

Friction along bimaterial interfaces is, perhaps, the most general mode of frictional (and tectonic) behavior. It is, however, probably the least studied.  The present study and the two previous ones~\cite{shlomai_PNAS,shlomai_jgr} allow us to draw a clear picture of bimaterial frictional rupture. The agreement between experiments, numerics and theory extends the fracture approach to this problem, that has been extremely successful in describing frictional ruptures within homogeneous interfaces. This study, together with  ~\cite{shlomai_PNAS,shlomai_jgr}, provides the nearly complete description of bimaterial rupture that is summarized below:

\begin{itemize}[leftmargin=*]

\item {\it Subsonic rupture, positive direction}. Crack-like propagation is stable up to $c_T$. As beyond $c_T$ cracks can not exist, the ruptures will transition to slip-pulses. 

\item Slip pulses occur only in the positive direction for $c>c_T$. They have a subsonic  limiting speed that merges with the generalized Rayleigh wave speed $c_{GR}$, for moderate bimaterial contrast and is transonic otherwise.

\item {\it Subsonic rupture, negative direction}. No other subsonic propagating modes exist, so $c_T$ is a limiting speed.

\item {\it Supershear rupture in the positive direction} are experimentally observed. They are rare, as they necessitate a large quantity of stored elastic energy prior to rupture nucleation. These ruptures are not a `pure' rupture mode but are composed of trains of slip pulses. This experimental observation is not described by theory, however the theory does predict that a {\it pure} supershear rupture mode can {\it not} exist in the positive direction. We believe that these trains of slip pulses may be  related to the nonlocality of the dissipation mechanism.

\item {\it Supershear ruptures in the negative direction} constitute the  most common rupture mode in the negative direction. These are possible in a finite range of supersonic speeds, whose existence is predicted by fracture theory ~\cite{shlomai_jgr}. The allowed supershear speeds, for the materials considered in Fig.~\ref{fig:Cs}, were provided in ~\cite{shlomai_jgr}.

\item {\it The different critical speeds} are predicted by fracture theory. $c_T$ exists for any bimaterial mismatch and satisfies $c_{R1}<c_T<c_{s1}$. $c^{lim}$ merges with $c_{GR}$ for low material contrast and is transonic, converging to $\sqrt{2}c_{s1}$, for infinite material contrast. This limiting case may be related to Schallamach waves~\cite{schallamach1971does,baumberger2002self}. These results are sketched in Fig.~\ref{fig:Cs} for model bimaterials that are characterized by a single material contrast parameter. 

\end{itemize}

While there is still more to be done, the studies described above have set the path. From the theoretical side, we need to understand the dissipative mechanisms;  when we are allowed to separate them and when we cannot. We believe that such a study will necessitate going beyond near-tip asymptotics and, therefore, will necessitate study of the whole elastodynamic problem. From the numerical side, we need to be guided by available experiments; positive vs negative directions, how to measure regions of dissipation and scale separation. More experiments are needed  - as only experiments will be able to discriminate between  different friction or dissipative laws at the interface. First steps have now been taken in experimental studies of  cohesive zone properties in homogeneous interfaces~\cite{Berman_2020}. Extending such studies to bimaterial interfaces could provide critical insights for these questions. On the experimental side, we also need data for more bimaterial contrasts, especially for the case when $c_{GR}$ exists in order to validate, for example, the predictions of Fig.~\ref{fig:Cs}. 

\begin{acknowledgments}

This work was supported by the International Research Project ``Non-Equilibrium Physics of Complex Systems'' (IRP-PhyComSys, France-Israel) and by the Laboratoire International Associ\'e ``Mati\`ere: Structure et Dynamique'' (LIA-MSD, France-Chile). M. A.-B. acknowledges the support of the Lady Davis Fellowship Trust. J. F. and H. S. acknowledge the support of the Israel Science Foundation (Grant 840/19).

\end{acknowledgments}

\bibliography{references}

\begin{thebibliography}{67}%
\makeatletter
\providecommand \@ifxundefined [1]{%
 \@ifx{#1\undefined}
}%
\providecommand \@ifnum [1]{%
 \ifnum #1\expandafter \@firstoftwo
 \else \expandafter \@secondoftwo
 \fi
}%
\providecommand \@ifx [1]{%
 \ifx #1\expandafter \@firstoftwo
 \else \expandafter \@secondoftwo
 \fi
}%
\providecommand \natexlab [1]{#1}%
\providecommand \enquote  [1]{``#1''}%
\providecommand \bibnamefont  [1]{#1}%
\providecommand \bibfnamefont [1]{#1}%
\providecommand \citenamefont [1]{#1}%
\providecommand \href@noop [0]{\@secondoftwo}%
\providecommand \href [0]{\begingroup \@sanitize@url \@href}%
\providecommand \@href[1]{\@@startlink{#1}\@@href}%
\providecommand \@@href[1]{\endgroup#1\@@endlink}%
\providecommand \@sanitize@url [0]{\catcode `\\12\catcode `\$12\catcode
  `\&12\catcode `\#12\catcode `\^12\catcode `\_12\catcode `\%12\relax}%
\providecommand \@@startlink[1]{}%
\providecommand \@@endlink[0]{}%
\providecommand \url  [0]{\begingroup\@sanitize@url \@url }%
\providecommand \@url [1]{\endgroup\@href {#1}{\urlprefix }}%
\providecommand \urlprefix  [0]{URL }%
\providecommand \Eprint [0]{\href }%
\providecommand \doibase [0]{http://dx.doi.org/}%
\providecommand \selectlanguage [0]{\@gobble}%
\providecommand \bibinfo  [0]{\@secondoftwo}%
\providecommand \bibfield  [0]{\@secondoftwo}%
\providecommand \translation [1]{[#1]}%
\providecommand \BibitemOpen [0]{}%
\providecommand \bibitemStop [0]{}%
\providecommand \bibitemNoStop [0]{.\EOS\space}%
\providecommand \EOS [0]{\spacefactor3000\relax}%
\providecommand \BibitemShut  [1]{\csname bibitem#1\endcsname}%
\let\auto@bib@innerbib\@empty
%</preamble>
\bibitem [{\citenamefont {Bowden}\ and\ \citenamefont
  {Tabor}(2001)}]{bowden2001friction}%
  \BibitemOpen
  \bibfield  {author} {\bibinfo {author} {\bibfnamefont {F.~P.}\ \bibnamefont
  {Bowden}}\ and\ \bibinfo {author} {\bibfnamefont {D.}~\bibnamefont {Tabor}},\
  }\href@noop {} {\emph {\bibinfo {title} {The Friction and Lubrication of
  Solids}}}\ (\bibinfo  {publisher} {Oxford University Press},\ \bibinfo {year}
  {2001})\BibitemShut {NoStop}%
\bibitem [{\citenamefont {Deresiewicz}(1988)}]{Deresiewicz1988}%
  \BibitemOpen
  \bibfield  {author} {\bibinfo {author} {\bibfnamefont {H.}~\bibnamefont
  {Deresiewicz}},\ }\bibfield  {title} {\enquote {\bibinfo {title} {Amontons
  and coulomb, friction's founding fathers},}\ }in\ \href@noop {} {\emph
  {\bibinfo {booktitle} {Approaches to Modeling of Friction and Wear}}}\
  (\bibinfo  {publisher} {Springer},\ \bibinfo {address} {New York},\ \bibinfo
  {year} {1988})\ pp.\ \bibinfo {pages} {56--60}\BibitemShut {NoStop}%
\bibitem [{\citenamefont {Dieterich}(1979)}]{dieterich1979modeling}%
  \BibitemOpen
  \bibfield  {author} {\bibinfo {author} {\bibfnamefont {J.~H}\ \bibnamefont
  {Dieterich}},\ }\bibfield  {title} {\enquote {\bibinfo {title} {Modeling of
  rock friction: 1. experimental results and constitutive equations},}\
  }\href@noop {} {\bibfield  {journal} {\bibinfo  {journal} {Journal of
  Geophysical Research: Solid Earth}\ }\textbf {\bibinfo {volume} {84}},\
  \bibinfo {pages} {2161--2168} (\bibinfo {year} {1979})}\BibitemShut {NoStop}%
\bibitem [{\citenamefont {Ruina}(1983)}]{ruina1983slip}%
  \BibitemOpen
  \bibfield  {author} {\bibinfo {author} {\bibfnamefont {A.}~\bibnamefont
  {Ruina}},\ }\bibfield  {title} {\enquote {\bibinfo {title} {Slip instability
  and state variable friction laws},}\ }\href@noop {} {\bibfield  {journal}
  {\bibinfo  {journal} {Journal of Geophysical Research: Solid Earth}\ }\textbf
  {\bibinfo {volume} {88}},\ \bibinfo {pages} {10359--10370} (\bibinfo {year}
  {1983})}\BibitemShut {NoStop}%
\bibitem [{\citenamefont {Rubinstein}\ \emph {et~al.}(2004)\citenamefont
  {Rubinstein}, \citenamefont {Cohen},\ and\ \citenamefont
  {Fineberg}}]{rubinstein2004detachment}%
  \BibitemOpen
  \bibfield  {author} {\bibinfo {author} {\bibfnamefont {S.~M.}\ \bibnamefont
  {Rubinstein}}, \bibinfo {author} {\bibfnamefont {G.}~\bibnamefont {Cohen}}, \
  and\ \bibinfo {author} {\bibfnamefont {J.}~\bibnamefont {Fineberg}},\
  }\bibfield  {title} {\enquote {\bibinfo {title} {Detachment fronts and the
  onset of dynamic friction},}\ }\href@noop {} {\bibfield  {journal} {\bibinfo
  {journal} {Nature}\ }\textbf {\bibinfo {volume} {430}},\ \bibinfo {pages}
  {1005--1009} (\bibinfo {year} {2004})}\BibitemShut {NoStop}%
\bibitem [{\citenamefont {Ben-David}\ \emph
  {et~al.}(2010{\natexlab{a}})\citenamefont {Ben-David}, \citenamefont
  {Cohen},\ and\ \citenamefont {Fineberg}}]{ben2010dynamics}%
  \BibitemOpen
  \bibfield  {author} {\bibinfo {author} {\bibfnamefont {O.}~\bibnamefont
  {Ben-David}}, \bibinfo {author} {\bibfnamefont {G.}~\bibnamefont {Cohen}}, \
  and\ \bibinfo {author} {\bibfnamefont {J.}~\bibnamefont {Fineberg}},\
  }\bibfield  {title} {\enquote {\bibinfo {title} {The dynamics of the onset of
  frictional slip},}\ }\href@noop {} {\bibfield  {journal} {\bibinfo  {journal}
  {Science}\ }\textbf {\bibinfo {volume} {330}},\ \bibinfo {pages} {211--214}
  (\bibinfo {year} {2010}{\natexlab{a}})}\BibitemShut {NoStop}%
\bibitem [{\citenamefont {Ben-David}\ \emph
  {et~al.}(2010{\natexlab{b}})\citenamefont {Ben-David}, \citenamefont
  {Rubinstein},\ and\ \citenamefont {Fineberg}}]{Ben-David2010Nature}%
  \BibitemOpen
  \bibfield  {author} {\bibinfo {author} {\bibfnamefont {O.}~\bibnamefont
  {Ben-David}}, \bibinfo {author} {\bibfnamefont {S.~M.}\ \bibnamefont
  {Rubinstein}}, \ and\ \bibinfo {author} {\bibfnamefont {J.}~\bibnamefont
  {Fineberg}},\ }\bibfield  {title} {\enquote {\bibinfo {title} {Slip-stick and
  the evolution of frictional strength},}\ }\href@noop {} {\bibfield  {journal}
  {\bibinfo  {journal} {Nature}\ }\textbf {\bibinfo {volume} {463}},\ \bibinfo
  {pages} {76--79} (\bibinfo {year} {2010}{\natexlab{b}})}\BibitemShut
  {NoStop}%
\bibitem [{\citenamefont {Ben-David}\ and\ \citenamefont
  {Fineberg}(2011)}]{Ben-David2011}%
  \BibitemOpen
  \bibfield  {author} {\bibinfo {author} {\bibfnamefont {O.}~\bibnamefont
  {Ben-David}}\ and\ \bibinfo {author} {\bibfnamefont {J.}~\bibnamefont
  {Fineberg}},\ }\bibfield  {title} {\enquote {\bibinfo {title} {Static
  friction coefficient is not a material constant},}\ }\href {\doibase
  10.1103/PhysRevLett.106.254301} {\bibfield  {journal} {\bibinfo  {journal}
  {Physical Review Letters}\ }\textbf {\bibinfo {volume} {106}},\ \bibinfo
  {pages} {254301} (\bibinfo {year} {2011})}\BibitemShut {NoStop}%
\bibitem [{\citenamefont {Svetlizky}\ and\ \citenamefont
  {Fineberg}(2014)}]{Svetlizky2014}%
  \BibitemOpen
  \bibfield  {author} {\bibinfo {author} {\bibfnamefont {I.}~\bibnamefont
  {Svetlizky}}\ and\ \bibinfo {author} {\bibfnamefont {J.}~\bibnamefont
  {Fineberg}},\ }\bibfield  {title} {\enquote {\bibinfo {title} {Classical
  shear cracks drive the onset of dry frictional motion},}\ }\href@noop {}
  {\bibfield  {journal} {\bibinfo  {journal} {Nature}\ }\textbf {\bibinfo
  {volume} {509}},\ \bibinfo {pages} {205--208} (\bibinfo {year}
  {2014})}\BibitemShut {NoStop}%
\bibitem [{\citenamefont {Shlomai}\ and\ \citenamefont
  {Fineberg}(2016)}]{shlomai2016structure}%
  \BibitemOpen
  \bibfield  {author} {\bibinfo {author} {\bibfnamefont {H.}~\bibnamefont
  {Shlomai}}\ and\ \bibinfo {author} {\bibfnamefont {J.}~\bibnamefont
  {Fineberg}},\ }\bibfield  {title} {\enquote {\bibinfo {title} {The structure
  of slip-pulses and supershear ruptures driving slip in bimaterial
  friction},}\ }\href@noop {} {\bibfield  {journal} {\bibinfo  {journal}
  {Nature communications}\ }\textbf {\bibinfo {volume} {7}},\ \bibinfo {pages}
  {1--7} (\bibinfo {year} {2016})}\BibitemShut {NoStop}%
\bibitem [{\citenamefont {Bayart}\ \emph {et~al.}(2016)\citenamefont {Bayart},
  \citenamefont {Svetlizky},\ and\ \citenamefont {Fineberg}}]{Bayart2016a}%
  \BibitemOpen
  \bibfield  {author} {\bibinfo {author} {\bibfnamefont {E.}~\bibnamefont
  {Bayart}}, \bibinfo {author} {\bibfnamefont {I.}~\bibnamefont {Svetlizky}}, \
  and\ \bibinfo {author} {\bibfnamefont {J.}~\bibnamefont {Fineberg}},\
  }\bibfield  {title} {\enquote {\bibinfo {title} {Fracture mechanics determine
  the lengths of interface ruptures that mediate frictional motion},}\
  }\href@noop {} {\bibfield  {journal} {\bibinfo  {journal} {Nature Physics}\
  }\textbf {\bibinfo {volume} {12}},\ \bibinfo {pages} {166--170} (\bibinfo
  {year} {2016})}\BibitemShut {NoStop}%
\bibitem [{\citenamefont {Svetlizky}\ \emph {et~al.}(2017)\citenamefont
  {Svetlizky}, \citenamefont {Kammer}, \citenamefont {Bayart}, \citenamefont
  {Cohen},\ and\ \citenamefont {Fineberg}}]{Svetlizky2017a}%
  \BibitemOpen
  \bibfield  {author} {\bibinfo {author} {\bibfnamefont {I.}~\bibnamefont
  {Svetlizky}}, \bibinfo {author} {\bibfnamefont {D.~S.}\ \bibnamefont
  {Kammer}}, \bibinfo {author} {\bibfnamefont {E.}~\bibnamefont {Bayart}},
  \bibinfo {author} {\bibfnamefont {G.}~\bibnamefont {Cohen}}, \ and\ \bibinfo
  {author} {\bibfnamefont {J.}~\bibnamefont {Fineberg}},\ }\bibfield  {title}
  {\enquote {\bibinfo {title} {Brittle fracture theory predicts the equation of
  motion of frictional rupture fronts},}\ }\href {\doibase
  10.1103/PhysRevLett.118.125501} {\bibfield  {journal} {\bibinfo  {journal}
  {Physical Review Letters}\ }\textbf {\bibinfo {volume} {118}},\ \bibinfo
  {pages} {125501} (\bibinfo {year} {2017})}\BibitemShut {NoStop}%
\bibitem [{\citenamefont {Bayart}\ \emph {et~al.}({2018})\citenamefont
  {Bayart}, \citenamefont {Svetlizky},\ and\ \citenamefont
  {Fineberg}}]{Bayart2019}%
  \BibitemOpen
  \bibfield  {author} {\bibinfo {author} {\bibfnamefont {E.}~\bibnamefont
  {Bayart}}, \bibinfo {author} {\bibfnamefont {I.}~\bibnamefont {Svetlizky}}, \
  and\ \bibinfo {author} {\bibfnamefont {J.}~\bibnamefont {Fineberg}},\
  }\bibfield  {title} {\enquote {\bibinfo {title} {{Rupture Dynamics of
  Heterogeneous Frictional Interfaces}},}\ }\href {\doibase
  {10.1002/2018JB015509}} {\bibfield  {journal} {\bibinfo  {journal} {{Journal
  os Geophysical Research: Solid Earth}}\ }\textbf {\bibinfo {volume}
  {{123}}},\ \bibinfo {pages} {{3828--3848}} (\bibinfo {year}
  {{2018}})}\BibitemShut {NoStop}%
\bibitem [{\citenamefont {Svetlizky}\ \emph {et~al.}(2020)\citenamefont
  {Svetlizky}, \citenamefont {Albertini}, \citenamefont {Cohen}, \citenamefont
  {Kammer},\ and\ \citenamefont {Fineberg}}]{Svetlizky2020}%
  \BibitemOpen
  \bibfield  {author} {\bibinfo {author} {\bibfnamefont {I.}~\bibnamefont
  {Svetlizky}}, \bibinfo {author} {\bibfnamefont {G.}~\bibnamefont
  {Albertini}}, \bibinfo {author} {\bibfnamefont {G.}~\bibnamefont {Cohen}},
  \bibinfo {author} {\bibfnamefont {D.~S.}\ \bibnamefont {Kammer}}, \ and\
  \bibinfo {author} {\bibfnamefont {J.}~\bibnamefont {Fineberg}},\ }\bibfield
  {title} {\enquote {\bibinfo {title} {Dynamic fields at the tip of
  sub-rayleigh and supershear frictional rupture fronts},}\ }\href {\doibase
  10.1016/j.jmps.2019.103826} {\bibfield  {journal} {\bibinfo  {journal}
  {Journal of the Mechanics and Physics of Solids}\ }\textbf {\bibinfo {volume}
  {137}},\ \bibinfo {pages} {103826} (\bibinfo {year} {2020})}\BibitemShut
  {NoStop}%
\bibitem [{\citenamefont {Shlomai}\ \emph
  {et~al.}(2020{\natexlab{a}})\citenamefont {Shlomai}, \citenamefont {Kammer},
  \citenamefont {Adda-Bedia},\ and\ \citenamefont {Fineberg}}]{shlomai_PNAS}%
  \BibitemOpen
  \bibfield  {author} {\bibinfo {author} {\bibfnamefont {H.}~\bibnamefont
  {Shlomai}}, \bibinfo {author} {\bibfnamefont {D.~S.}\ \bibnamefont {Kammer}},
  \bibinfo {author} {\bibfnamefont {M.}~\bibnamefont {Adda-Bedia}}, \ and\
  \bibinfo {author} {\bibfnamefont {J.}~\bibnamefont {Fineberg}},\ }\bibfield
  {title} {\enquote {\bibinfo {title} {The onset of the frictional motion of
  dissimilar materials},}\ }\href@noop {} {\bibfield  {journal} {\bibinfo
  {journal} {Proceedings of the National Academy of Sciences}\ }\textbf
  {\bibinfo {volume} {117}},\ \bibinfo {pages} {13914--13920} (\bibinfo {year}
  {2020}{\natexlab{a}})}\BibitemShut {NoStop}%
\bibitem [{\citenamefont {Shlomai}\ \emph
  {et~al.}(2020{\natexlab{b}})\citenamefont {Shlomai}, \citenamefont
  {Adda-Bedia}, \citenamefont {Arias},\ and\ \citenamefont
  {Fineberg}}]{shlomai_jgr}%
  \BibitemOpen
  \bibfield  {author} {\bibinfo {author} {\bibfnamefont {H.}~\bibnamefont
  {Shlomai}}, \bibinfo {author} {\bibfnamefont {M.}~\bibnamefont {Adda-Bedia}},
  \bibinfo {author} {\bibfnamefont {R.}~\bibnamefont {Arias}}, \ and\ \bibinfo
  {author} {\bibfnamefont {J.}~\bibnamefont {Fineberg}},\ }\bibfield  {title}
  {\enquote {\bibinfo {title} {Supershear frictional crack along bimaterial
  interfaces},}\ }\href@noop {} {\bibfield  {journal} {\bibinfo  {journal}
  {Journal of Geophysical Research: Solid Earth}\ }\textbf {\bibinfo {volume}
  {125}},\ \bibinfo {pages} {e2020JB019829} (\bibinfo {year}
  {2020}{\natexlab{b}})}\BibitemShut {NoStop}%
\bibitem [{\citenamefont {Rosakis}(2002)}]{rosakis2002intersonic}%
  \BibitemOpen
  \bibfield  {author} {\bibinfo {author} {\bibfnamefont {A.~J.}\ \bibnamefont
  {Rosakis}},\ }\bibfield  {title} {\enquote {\bibinfo {title} {Intersonic
  shear cracks and fault ruptures},}\ }\href@noop {} {\bibfield  {journal}
  {\bibinfo  {journal} {Advances in Physics}\ }\textbf {\bibinfo {volume}
  {51}},\ \bibinfo {pages} {1189--1257} (\bibinfo {year} {2002})}\BibitemShut
  {NoStop}%
\bibitem [{\citenamefont {Ohnaka}\ and\ \citenamefont
  {Shen}(1999)}]{ohnaka1999scaling}%
  \BibitemOpen
  \bibfield  {author} {\bibinfo {author} {\bibfnamefont {M.}~\bibnamefont
  {Ohnaka}}\ and\ \bibinfo {author} {\bibfnamefont {L.-F.}\ \bibnamefont
  {Shen}},\ }\bibfield  {title} {\enquote {\bibinfo {title} {Scaling of the
  shear rupture process from nucleation to dynamic propagation: Implications of
  geometric irregularity of the rupturing surfaces},}\ }\href@noop {}
  {\bibfield  {journal} {\bibinfo  {journal} {Journal of Geophysical Research:
  Solid Earth}\ }\textbf {\bibinfo {volume} {104}},\ \bibinfo {pages}
  {817--844} (\bibinfo {year} {1999})}\BibitemShut {NoStop}%
\bibitem [{\citenamefont {Xu}\ \emph {et~al.}(2018)\citenamefont {Xu},
  \citenamefont {Fukuyama}, \citenamefont {Yamashita}, \citenamefont
  {Mizoguchi}, \citenamefont {Takizawa},\ and\ \citenamefont
  {Kawakata}}]{Xu_Fukuyama_2018}%
  \BibitemOpen
  \bibfield  {author} {\bibinfo {author} {\bibfnamefont {S.}~\bibnamefont
  {Xu}}, \bibinfo {author} {\bibfnamefont {E.}~\bibnamefont {Fukuyama}},
  \bibinfo {author} {\bibfnamefont {F.}~\bibnamefont {Yamashita}}, \bibinfo
  {author} {\bibfnamefont {K.}~\bibnamefont {Mizoguchi}}, \bibinfo {author}
  {\bibfnamefont {S.}~\bibnamefont {Takizawa}}, \ and\ \bibinfo {author}
  {\bibfnamefont {H.}~\bibnamefont {Kawakata}},\ }\bibfield  {title} {\enquote
  {\bibinfo {title} {Strain rate effect on fault slip and rupture evolution:
  Insight from meter-scale rock friction experiments},}\ }\href {\doibase
  10.1016/j.tecto.2017.11.039} {\bibfield  {journal} {\bibinfo  {journal}
  {Tectonophysics}\ }\textbf {\bibinfo {volume} {733}},\ \bibinfo {pages}
  {209--231} (\bibinfo {year} {2018})}\BibitemShut {NoStop}%
\bibitem [{\citenamefont {Kammer}\ \emph {et~al.}(2014)\citenamefont {Kammer},
  \citenamefont {Yastrebov}, \citenamefont {Anciaux},\ and\ \citenamefont
  {Molinari}}]{kammer2014existence}%
  \BibitemOpen
  \bibfield  {author} {\bibinfo {author} {\bibfnamefont {D.~S.}\ \bibnamefont
  {Kammer}}, \bibinfo {author} {\bibfnamefont {V.~A.}\ \bibnamefont
  {Yastrebov}}, \bibinfo {author} {\bibfnamefont {G.}~\bibnamefont {Anciaux}},
  \ and\ \bibinfo {author} {\bibfnamefont {J.-F.}\ \bibnamefont {Molinari}},\
  }\bibfield  {title} {\enquote {\bibinfo {title} {The existence of a critical
  length scale in regularised friction},}\ }\href@noop {} {\bibfield  {journal}
  {\bibinfo  {journal} {Journal of the Mechanics and Physics of Solids}\
  }\textbf {\bibinfo {volume} {63}},\ \bibinfo {pages} {40--50} (\bibinfo
  {year} {2014})}\BibitemShut {NoStop}%
\bibitem [{\citenamefont {Freund}(1998)}]{freund1998dynamic}%
  \BibitemOpen
  \bibfield  {author} {\bibinfo {author} {\bibfnamefont {L.~B.}\ \bibnamefont
  {Freund}},\ }\href@noop {} {\emph {\bibinfo {title} {Dynamic Fracture
  Mechanics}}}\ (\bibinfo  {publisher} {Cambridge University Press},\ \bibinfo
  {year} {1998})\BibitemShut {NoStop}%
\bibitem [{\citenamefont {Barras}\ \emph {et~al.}(2020)\citenamefont {Barras},
  \citenamefont {Aldam}, \citenamefont {Roch}, \citenamefont {Brener},
  \citenamefont {Bouchbinder},\ and\ \citenamefont {Molinari}}]{Barras_2020}%
  \BibitemOpen
  \bibfield  {author} {\bibinfo {author} {\bibfnamefont {F.}~\bibnamefont
  {Barras}}, \bibinfo {author} {\bibfnamefont {M.}~\bibnamefont {Aldam}},
  \bibinfo {author} {\bibfnamefont {T.}~\bibnamefont {Roch}}, \bibinfo {author}
  {\bibfnamefont {E.~A.}\ \bibnamefont {Brener}}, \bibinfo {author}
  {\bibfnamefont {E.}~\bibnamefont {Bouchbinder}}, \ and\ \bibinfo {author}
  {\bibfnamefont {J.-F.}\ \bibnamefont {Molinari}},\ }\bibfield  {title}
  {\enquote {\bibinfo {title} {The emergence of crack-like behavior of
  frictional rupture: Edge singularity and energy balance},}\ }\href@noop {}
  {\bibfield  {journal} {\bibinfo  {journal} {Earth and Planetray Science
  Letters}\ }\textbf {\bibinfo {volume} {531}} (\bibinfo {year}
  {2020})}\BibitemShut {NoStop}%
\bibitem [{\citenamefont {Palmer}\ and\ \citenamefont
  {Rice}(1973)}]{palmer1973growth}%
  \BibitemOpen
  \bibfield  {author} {\bibinfo {author} {\bibfnamefont {A.~C.}\ \bibnamefont
  {Palmer}}\ and\ \bibinfo {author} {\bibfnamefont {J.~R.}\ \bibnamefont
  {Rice}},\ }\bibfield  {title} {\enquote {\bibinfo {title} {The growth of slip
  surfaces in the progressive failure of over-consolidated clay},}\ }\href@noop
  {} {\bibfield  {journal} {\bibinfo  {journal} {Proceedings of the Royal
  Society of London. Series A: Mathematical and Physical Sciences}\ }\textbf
  {\bibinfo {volume} {332}},\ \bibinfo {pages} {527--548} (\bibinfo {year}
  {1973})}\BibitemShut {NoStop}%
\bibitem [{\citenamefont {Das}(2003)}]{das2003dynamic}%
  \BibitemOpen
  \bibfield  {author} {\bibinfo {author} {\bibfnamefont {S.}~\bibnamefont
  {Das}},\ }\bibfield  {title} {\enquote {\bibinfo {title} {Dynamic fracture
  mechanics in the study of the earthquake rupturing process: theory and
  observation},}\ }\href@noop {} {\bibfield  {journal} {\bibinfo  {journal}
  {Journal of the Mechanics and Physics of Solids}\ }\textbf {\bibinfo {volume}
  {51}},\ \bibinfo {pages} {1939--1955} (\bibinfo {year} {2003})}\BibitemShut
  {NoStop}%
\bibitem [{\citenamefont {Aldam}\ \emph {et~al.}(2016)\citenamefont {Aldam},
  \citenamefont {Bar-Sinai}, \citenamefont {Svetlizky}, \citenamefont {Brener},
  \citenamefont {Fineberg},\ and\ \citenamefont
  {Bouchbinder}}]{AldamBouchbinder2016GeometricBimaterial}%
  \BibitemOpen
  \bibfield  {author} {\bibinfo {author} {\bibfnamefont {M.}~\bibnamefont
  {Aldam}}, \bibinfo {author} {\bibfnamefont {Y.}~\bibnamefont {Bar-Sinai}},
  \bibinfo {author} {\bibfnamefont {I.}~\bibnamefont {Svetlizky}}, \bibinfo
  {author} {\bibfnamefont {E.~A.}\ \bibnamefont {Brener}}, \bibinfo {author}
  {\bibfnamefont {J.}~\bibnamefont {Fineberg}}, \ and\ \bibinfo {author}
  {\bibfnamefont {E.}~\bibnamefont {Bouchbinder}},\ }\bibfield  {title}
  {\enquote {\bibinfo {title} {Frictional sliding without geometrical
  reflection symmetry},}\ }\href@noop {} {\bibfield  {journal} {\bibinfo
  {journal} {Physical Review X}\ }\textbf {\bibinfo {volume} {6}},\ \bibinfo
  {pages} {041023} (\bibinfo {year} {2016})}\BibitemShut {NoStop}%
\bibitem [{\citenamefont {Weertman}(1963)}]{weertman1963dislocations}%
  \BibitemOpen
  \bibfield  {author} {\bibinfo {author} {\bibfnamefont {J.}~\bibnamefont
  {Weertman}},\ }\bibfield  {title} {\enquote {\bibinfo {title} {Dislocations
  moving uniformly on the interface between isotropic media of different
  elastic properties},}\ }\href@noop {} {\bibfield  {journal} {\bibinfo
  {journal} {Journal of the Mechanics and Physics of Solids}\ }\textbf
  {\bibinfo {volume} {11}},\ \bibinfo {pages} {197--204} (\bibinfo {year}
  {1963})}\BibitemShut {NoStop}%
\bibitem [{\citenamefont {Weertman}(1980)}]{weertman1980unstable}%
  \BibitemOpen
  \bibfield  {author} {\bibinfo {author} {\bibfnamefont {J.}~\bibnamefont
  {Weertman}},\ }\bibfield  {title} {\enquote {\bibinfo {title} {Unstable
  slippage across a fault that separates elastic media of different elastic
  constants},}\ }\href@noop {} {\bibfield  {journal} {\bibinfo  {journal}
  {Journal of Geophysical Research: Solid Earth}\ }\textbf {\bibinfo {volume}
  {85}},\ \bibinfo {pages} {1455--1461} (\bibinfo {year} {1980})}\BibitemShut
  {NoStop}%
\bibitem [{\citenamefont {Ampuero}\ and\ \citenamefont
  {Ben-Zion}(2008)}]{AmpueroBenZion2008}%
  \BibitemOpen
  \bibfield  {author} {\bibinfo {author} {\bibfnamefont {J.-P.}\ \bibnamefont
  {Ampuero}}\ and\ \bibinfo {author} {\bibfnamefont {Y.}~\bibnamefont
  {Ben-Zion}},\ }\bibfield  {title} {\enquote {\bibinfo {title} {Cracks, pulses
  and macroscopic asymmetry of dynamic rupture on a bimaterial interface with
  velocity-weakening friction},}\ }\href {\doibase
  10.1111/j.1365-246X.2008.03736.x} {\bibfield  {journal} {\bibinfo  {journal}
  {Geophysical International Journal}\ }\textbf {\bibinfo {volume} {173}},\
  \bibinfo {pages} {674--692} (\bibinfo {year} {2008})}\BibitemShut {NoStop}%
\bibitem [{\citenamefont {Scala}\ \emph {et~al.}(2017)\citenamefont {Scala},
  \citenamefont {Festa},\ and\ \citenamefont {Vilotte}}]{Scala2017}%
  \BibitemOpen
  \bibfield  {author} {\bibinfo {author} {\bibfnamefont {A.}~\bibnamefont
  {Scala}}, \bibinfo {author} {\bibfnamefont {G.}~\bibnamefont {Festa}}, \ and\
  \bibinfo {author} {\bibfnamefont {J.-P.}\ \bibnamefont {Vilotte}},\
  }\bibfield  {title} {\enquote {\bibinfo {title} {{Rupture dynamics along
  bimaterial interfaces: a parametric study of the shear-normal traction
  coupling}},}\ }\href@noop {} {\bibfield  {journal} {\bibinfo  {journal}
  {Geophysical Journal International}\ }\textbf {\bibinfo {volume} {209}},\
  \bibinfo {pages} {48--67} (\bibinfo {year} {2017})}\BibitemShut {NoStop}%
\bibitem [{\citenamefont {Perrin}\ \emph {et~al.}(1995)\citenamefont {Perrin},
  \citenamefont {Rice},\ and\ \citenamefont {Zheng}}]{perrin1995self}%
  \BibitemOpen
  \bibfield  {author} {\bibinfo {author} {\bibfnamefont {G.}~\bibnamefont
  {Perrin}}, \bibinfo {author} {\bibfnamefont {J.~R.}\ \bibnamefont {Rice}}, \
  and\ \bibinfo {author} {\bibfnamefont {G.}~\bibnamefont {Zheng}},\ }\bibfield
   {title} {\enquote {\bibinfo {title} {Self-healing slip pulse on a frictional
  surface},}\ }\href@noop {} {\bibfield  {journal} {\bibinfo  {journal}
  {Journal of the Mechanics and Physics of Solids}\ }\textbf {\bibinfo {volume}
  {43}},\ \bibinfo {pages} {1461--1495} (\bibinfo {year} {1995})}\BibitemShut
  {NoStop}%
\bibitem [{\citenamefont {Heimisson}\ \emph {et~al.}(2019)\citenamefont
  {Heimisson}, \citenamefont {Dunham},\ and\ \citenamefont
  {Almquist}}]{Heimisson2019}%
  \BibitemOpen
  \bibfield  {author} {\bibinfo {author} {\bibfnamefont {E.}~\bibnamefont
  {Heimisson}}, \bibinfo {author} {\bibfnamefont {E.}~\bibnamefont {Dunham}}, \
  and\ \bibinfo {author} {\bibfnamefont {M.}~\bibnamefont {Almquist}},\
  }\bibfield  {title} {\enquote {\bibinfo {title} {Poroelastic effects
  destabilize mildly rate-strengthening friction to generate stable slow slip
  pulses},}\ }\href@noop {} {\bibfield  {journal} {\bibinfo  {journal} {Journal
  of the Mechanics and Physics of Solids}\ }\textbf {\bibinfo {volume} {130}}
  (\bibinfo {year} {2019})}\BibitemShut {NoStop}%
\bibitem [{\citenamefont {Brantut}\ \emph {et~al.}(2019)\citenamefont
  {Brantut}, \citenamefont {Garagash},\ and\ \citenamefont
  {Noda}}]{brantut2019stability}%
  \BibitemOpen
  \bibfield  {author} {\bibinfo {author} {\bibfnamefont {N.}~\bibnamefont
  {Brantut}}, \bibinfo {author} {\bibfnamefont {D.~I.}\ \bibnamefont
  {Garagash}}, \ and\ \bibinfo {author} {\bibfnamefont {H.}~\bibnamefont
  {Noda}},\ }\bibfield  {title} {\enquote {\bibinfo {title} {Stability of
  pulse-like earthquake ruptures},}\ }\href@noop {} {\bibfield  {journal}
  {\bibinfo  {journal} {Journal of Geophysical Research: Solid Earth}\ }\textbf
  {\bibinfo {volume} {124}},\ \bibinfo {pages} {8998--9020} (\bibinfo {year}
  {2019})}\BibitemShut {NoStop}%
\bibitem [{\citenamefont {Xia}\ \emph {et~al.}(2005{\natexlab{a}})\citenamefont
  {Xia}, \citenamefont {Rosakis}, \citenamefont {Kanamori},\ and\ \citenamefont
  {Rice}}]{xia2005laboratory}%
  \BibitemOpen
  \bibfield  {author} {\bibinfo {author} {\bibfnamefont {K.}~\bibnamefont
  {Xia}}, \bibinfo {author} {\bibfnamefont {A.~J.}\ \bibnamefont {Rosakis}},
  \bibinfo {author} {\bibfnamefont {H.}~\bibnamefont {Kanamori}}, \ and\
  \bibinfo {author} {\bibfnamefont {J.~R.}\ \bibnamefont {Rice}},\ }\bibfield
  {title} {\enquote {\bibinfo {title} {Laboratory earthquakes along
  inhomogeneous faults: Directionality and supershear},}\ }\href@noop {}
  {\bibfield  {journal} {\bibinfo  {journal} {Science}\ }\textbf {\bibinfo
  {volume} {308}},\ \bibinfo {pages} {681--684} (\bibinfo {year}
  {2005}{\natexlab{a}})}\BibitemShut {NoStop}%
\bibitem [{\citenamefont {Lykotrafitis}\ \emph {et~al.}(2006)\citenamefont
  {Lykotrafitis}, \citenamefont {Rosakis},\ and\ \citenamefont
  {Ravichandran}}]{lykotrafitis2006self}%
  \BibitemOpen
  \bibfield  {author} {\bibinfo {author} {\bibfnamefont {G.}~\bibnamefont
  {Lykotrafitis}}, \bibinfo {author} {\bibfnamefont {A.~J.}\ \bibnamefont
  {Rosakis}}, \ and\ \bibinfo {author} {\bibfnamefont {G.}~\bibnamefont
  {Ravichandran}},\ }\bibfield  {title} {\enquote {\bibinfo {title}
  {Self-healing pulse-like shear ruptures in the laboratory},}\ }\href@noop {}
  {\bibfield  {journal} {\bibinfo  {journal} {Science}\ }\textbf {\bibinfo
  {volume} {313}},\ \bibinfo {pages} {1765--1768} (\bibinfo {year}
  {2006})}\BibitemShut {NoStop}%
\bibitem [{\citenamefont {Rayleigh}(1885)}]{rayleigh1885waves}%
  \BibitemOpen
  \bibfield  {author} {\bibinfo {author} {\bibfnamefont {L.}~\bibnamefont
  {Rayleigh}},\ }\bibfield  {title} {\enquote {\bibinfo {title} {On waves
  propagated along the plane surface of an elastic solid},}\ }\href@noop {}
  {\bibfield  {journal} {\bibinfo  {journal} {Proceedings of the London
  Mathematical Society}\ }\textbf {\bibinfo {volume} {1}},\ \bibinfo {pages}
  {4--11} (\bibinfo {year} {1885})}\BibitemShut {NoStop}%
\bibitem [{\citenamefont {Stroh}(1957)}]{stroh1957theory}%
  \BibitemOpen
  \bibfield  {author} {\bibinfo {author} {\bibfnamefont {A.~N.}\ \bibnamefont
  {Stroh}},\ }\bibfield  {title} {\enquote {\bibinfo {title} {A theory of the
  fracture of metals},}\ }\href@noop {} {\bibfield  {journal} {\bibinfo
  {journal} {Advances in Physics}\ }\textbf {\bibinfo {volume} {6}},\ \bibinfo
  {pages} {418--465} (\bibinfo {year} {1957})}\BibitemShut {NoStop}%
\bibitem [{\citenamefont {Stoneley}(1924)}]{stoneley1924elastic}%
  \BibitemOpen
  \bibfield  {author} {\bibinfo {author} {\bibfnamefont {R.}~\bibnamefont
  {Stoneley}},\ }\bibfield  {title} {\enquote {\bibinfo {title} {Elastic waves
  at the surface of separation of two solids},}\ }\href@noop {} {\bibfield
  {journal} {\bibinfo  {journal} {Proceedings of the Royal Society of London.
  Series A: Containing Papers of a Mathematical and Physical Character}\
  }\textbf {\bibinfo {volume} {106}},\ \bibinfo {pages} {416--428} (\bibinfo
  {year} {1924})}\BibitemShut {NoStop}%
\bibitem [{\citenamefont {Gol'dshtein}(1967)}]{gol1967surface}%
  \BibitemOpen
  \bibfield  {author} {\bibinfo {author} {\bibfnamefont {R.~V.}\ \bibnamefont
  {Gol'dshtein}},\ }\bibfield  {title} {\enquote {\bibinfo {title} {On surface
  waves in joined elastic materials and their relation to crack propagation
  along the junction},}\ }\href@noop {} {\bibfield  {journal} {\bibinfo
  {journal} {Journal of Applied Mathematics and Mechanics}\ }\textbf {\bibinfo
  {volume} {31}},\ \bibinfo {pages} {497--502} (\bibinfo {year}
  {1967})}\BibitemShut {NoStop}%
\bibitem [{\citenamefont {Adams}(1995)}]{adams1995self}%
  \BibitemOpen
  \bibfield  {author} {\bibinfo {author} {\bibfnamefont {G.~G.}\ \bibnamefont
  {Adams}},\ }\bibfield  {title} {\enquote {\bibinfo {title} {Self-excited
  oscillations of two elastic half-spaces sliding with a constant coefficient
  of friction},}\ }\href@noop {} {\bibfield  {journal} {\bibinfo  {journal}
  {Journal of Applied Mechanics}\ }\textbf {\bibinfo {volume} {62}},\ \bibinfo
  {pages} {867} (\bibinfo {year} {1995})}\BibitemShut {NoStop}%
\bibitem [{\citenamefont {Ranjith}\ and\ \citenamefont
  {Rice}(2001)}]{ranjith2001slip}%
  \BibitemOpen
  \bibfield  {author} {\bibinfo {author} {\bibfnamefont {K.}~\bibnamefont
  {Ranjith}}\ and\ \bibinfo {author} {\bibfnamefont {J.~R.}\ \bibnamefont
  {Rice}},\ }\bibfield  {title} {\enquote {\bibinfo {title} {Slip dynamics at
  an interface between dissimilar materials},}\ }\href@noop {} {\bibfield
  {journal} {\bibinfo  {journal} {Journal of the Mechanics and Physics of
  Solids}\ }\textbf {\bibinfo {volume} {49}},\ \bibinfo {pages} {341--361}
  (\bibinfo {year} {2001})}\BibitemShut {NoStop}%
\bibitem [{\citenamefont {Willis}(1971)}]{willis1971fracture}%
  \BibitemOpen
  \bibfield  {author} {\bibinfo {author} {\bibfnamefont {J.~R.}\ \bibnamefont
  {Willis}},\ }\bibfield  {title} {\enquote {\bibinfo {title} {Fracture
  mechanics of interfacial cracks},}\ }\href@noop {} {\bibfield  {journal}
  {\bibinfo  {journal} {Journal of the Mechanics and Physics of Solids}\
  }\textbf {\bibinfo {volume} {19}},\ \bibinfo {pages} {353--368} (\bibinfo
  {year} {1971})}\BibitemShut {NoStop}%
\bibitem [{\citenamefont {Atkinson}(1977)}]{atkinson1977dynamic}%
  \BibitemOpen
  \bibfield  {author} {\bibinfo {author} {\bibfnamefont {C.}~\bibnamefont
  {Atkinson}},\ }\bibfield  {title} {\enquote {\bibinfo {title} {Dynamic crack
  problems in dissimilar media},}\ }in\ \href@noop {} {\emph {\bibinfo
  {booktitle} {Mechanics of Fracture}}},\ Vol.~\bibinfo {volume} {4},\ \bibinfo
  {editor} {edited by\ \bibinfo {editor} {\bibfnamefont {G.~C.}\ \bibnamefont
  {Sih}}}\ (\bibinfo  {publisher} {Noordhoff, Leyden},\ \bibinfo {year}
  {1977})\ pp.\ \bibinfo {pages} {213--248}\BibitemShut {NoStop}%
\bibitem [{\citenamefont {Yang}\ \emph {et~al.}(1991)\citenamefont {Yang},
  \citenamefont {Suo},\ and\ \citenamefont {Shih}}]{yang1991mechanics}%
  \BibitemOpen
  \bibfield  {author} {\bibinfo {author} {\bibfnamefont {W.}~\bibnamefont
  {Yang}}, \bibinfo {author} {\bibfnamefont {Z.}~\bibnamefont {Suo}}, \ and\
  \bibinfo {author} {\bibfnamefont {C.~F.}\ \bibnamefont {Shih}},\ }\bibfield
  {title} {\enquote {\bibinfo {title} {Mechanics of dynamic debonding},}\
  }\href@noop {} {\bibfield  {journal} {\bibinfo  {journal} {Proceedings of the
  Royal Society of London. Series A: Mathematical and Physical Sciences}\
  }\textbf {\bibinfo {volume} {433}},\ \bibinfo {pages} {679--697} (\bibinfo
  {year} {1991})}\BibitemShut {NoStop}%
\bibitem [{\citenamefont {Liu}\ \emph {et~al.}(1993)\citenamefont {Liu},
  \citenamefont {Lambros},\ and\ \citenamefont {Rosakis}}]{liu1993highly}%
  \BibitemOpen
  \bibfield  {author} {\bibinfo {author} {\bibfnamefont {C.}~\bibnamefont
  {Liu}}, \bibinfo {author} {\bibfnamefont {J.}~\bibnamefont {Lambros}}, \ and\
  \bibinfo {author} {\bibfnamefont {A.~J.}\ \bibnamefont {Rosakis}},\
  }\bibfield  {title} {\enquote {\bibinfo {title} {Highly transient
  elastodynamic crack growth in a bimaterial interface: higher order asymptotic
  analysis and optical experiments},}\ }\href@noop {} {\bibfield  {journal}
  {\bibinfo  {journal} {Journal of the Mechanics and Physics of Solids}\
  }\textbf {\bibinfo {volume} {41}},\ \bibinfo {pages} {1887--1954} (\bibinfo
  {year} {1993})}\BibitemShut {NoStop}%
\bibitem [{\citenamefont {Lambros}\ and\ \citenamefont
  {Rosakis}(1995)}]{lambros1995shear}%
  \BibitemOpen
  \bibfield  {author} {\bibinfo {author} {\bibfnamefont {J.}~\bibnamefont
  {Lambros}}\ and\ \bibinfo {author} {\bibfnamefont {A.~J.}\ \bibnamefont
  {Rosakis}},\ }\bibfield  {title} {\enquote {\bibinfo {title} {Shear dominated
  transonic interfacial crack growth in a bimaterial-i. experimental
  observations},}\ }\href@noop {} {\bibfield  {journal} {\bibinfo  {journal}
  {Journal of the Mechanics and Physics of Solids}\ }\textbf {\bibinfo {volume}
  {43}},\ \bibinfo {pages} {169--188} (\bibinfo {year} {1995})}\BibitemShut
  {NoStop}%
\bibitem [{\citenamefont {Xia}\ \emph {et~al.}(2005{\natexlab{b}})\citenamefont
  {Xia}, \citenamefont {Rosakis},\ and\ \citenamefont
  {Kanamori}}]{XiaRosakis2005SupershearTransition}%
  \BibitemOpen
  \bibfield  {author} {\bibinfo {author} {\bibfnamefont {K.}~\bibnamefont
  {Xia}}, \bibinfo {author} {\bibfnamefont {A.~J.}\ \bibnamefont {Rosakis}}, \
  and\ \bibinfo {author} {\bibfnamefont {H.}~\bibnamefont {Kanamori}},\
  }\bibfield  {title} {\enquote {\bibinfo {title} {Supershear and subrayleigh
  to supershear transition observed in laboratory earthquake experiments},}\
  }\href {\doibase 10.1111/j.1747-1567.2005.tb00220.x} {\bibfield  {journal}
  {\bibinfo  {journal} {Experimental Techniques}\ }\textbf {\bibinfo {volume}
  {29}},\ \bibinfo {pages} {63--66} (\bibinfo {year}
  {2005}{\natexlab{b}})}\BibitemShut {NoStop}%
\bibitem [{\citenamefont {Rice}(1988)}]{Rice1988Elastic}%
  \BibitemOpen
  \bibfield  {author} {\bibinfo {author} {\bibfnamefont {J.~R.}\ \bibnamefont
  {Rice}},\ }\bibfield  {title} {\enquote {\bibinfo {title} {{Elastic Fracture
  Mechanics Concepts for Interfacial Cracks}},}\ }\href {\doibase
  10.1115/1.3173668} {\bibfield  {journal} {\bibinfo  {journal} {Journal of
  Applied Mechanics}\ }\textbf {\bibinfo {volume} {55}},\ \bibinfo {pages}
  {98--103} (\bibinfo {year} {1988})}\BibitemShut {NoStop}%
\bibitem [{\citenamefont {Deng}(1993)}]{deng1993propagating}%
  \BibitemOpen
  \bibfield  {author} {\bibinfo {author} {\bibfnamefont {X.}~\bibnamefont
  {Deng}},\ }\bibfield  {title} {\enquote {\bibinfo {title} {Propagating
  interface cracks with frictionless contact},}\ }\href@noop {} {\bibfield
  {journal} {\bibinfo  {journal} {Journal of the Mechanics and Physics of
  Solids}\ }\textbf {\bibinfo {volume} {41}},\ \bibinfo {pages} {531--540}
  (\bibinfo {year} {1993})}\BibitemShut {NoStop}%
\bibitem [{\citenamefont {Cochard}\ and\ \citenamefont
  {Rice}(2000)}]{cochard2000fault}%
  \BibitemOpen
  \bibfield  {author} {\bibinfo {author} {\bibfnamefont {A.}~\bibnamefont
  {Cochard}}\ and\ \bibinfo {author} {\bibfnamefont {J.~R.}\ \bibnamefont
  {Rice}},\ }\bibfield  {title} {\enquote {\bibinfo {title} {Fault rupture
  between dissimilar materials: Ill-posedness, regularization, and slip-pulse
  response},}\ }\href@noop {} {\bibfield  {journal} {\bibinfo  {journal}
  {Journal of Geophysical Research: Solid Earth}\ }\textbf {\bibinfo {volume}
  {105}},\ \bibinfo {pages} {25891--25907} (\bibinfo {year}
  {2000})}\BibitemShut {NoStop}%
\bibitem [{\citenamefont {Rice}(1980)}]{rice1980mechanics}%
  \BibitemOpen
  \bibfield  {author} {\bibinfo {author} {\bibfnamefont {J.~R.}\ \bibnamefont
  {Rice}},\ }\bibfield  {title} {\enquote {\bibinfo {title} {The mechanics of
  earthquake rupture},}\ }in\ \href@noop {} {\emph {\bibinfo {booktitle}
  {Physics of the Earth's Interior}}},\ \bibinfo {editor} {edited by\ \bibinfo
  {editor} {\bibfnamefont {A.~M.}\ \bibnamefont {Dziewonski}}\ and\ \bibinfo
  {editor} {\bibfnamefont {E.}~\bibnamefont {Boschi}}},\ \bibinfo
  {organization} {Italian Physical Society}\ (\bibinfo  {publisher} {North
  Holland},\ \bibinfo {year} {1980})\BibitemShut {NoStop}%
\bibitem [{\citenamefont {Adda-Bedia}\ \emph
  {et~al.}(1999{\natexlab{a}})\citenamefont {Adda-Bedia}, \citenamefont
  {Arias}, \citenamefont {Ben~Amar},\ and\ \citenamefont
  {Lund}}]{adda1999dynamic}%
  \BibitemOpen
  \bibfield  {author} {\bibinfo {author} {\bibfnamefont {M.}~\bibnamefont
  {Adda-Bedia}}, \bibinfo {author} {\bibfnamefont {R.}~\bibnamefont {Arias}},
  \bibinfo {author} {\bibfnamefont {M.}~\bibnamefont {Ben~Amar}}, \ and\
  \bibinfo {author} {\bibfnamefont {F.}~\bibnamefont {Lund}},\ }\bibfield
  {title} {\enquote {\bibinfo {title} {Dynamic instability of brittle
  fracture},}\ }\href@noop {} {\bibfield  {journal} {\bibinfo  {journal}
  {Physical Review Letters}\ }\textbf {\bibinfo {volume} {82}},\ \bibinfo
  {pages} {2314} (\bibinfo {year} {1999}{\natexlab{a}})}\BibitemShut {NoStop}%
\bibitem [{\citenamefont {Adda-Bedia}\ \emph
  {et~al.}(1999{\natexlab{b}})\citenamefont {Adda-Bedia}, \citenamefont
  {Arias}, \citenamefont {Ben~Amar},\ and\ \citenamefont {Lund}}]{Adda1999}%
  \BibitemOpen
  \bibfield  {author} {\bibinfo {author} {\bibfnamefont {M.}~\bibnamefont
  {Adda-Bedia}}, \bibinfo {author} {\bibfnamefont {R.}~\bibnamefont {Arias}},
  \bibinfo {author} {\bibfnamefont {M.}~\bibnamefont {Ben~Amar}}, \ and\
  \bibinfo {author} {\bibfnamefont {F.}~\bibnamefont {Lund}},\ }\bibfield
  {title} {\enquote {\bibinfo {title} {Generalized griffith criterion for
  dynamic fracture and the stability of crack motion at high velocities},}\
  }\href {\doibase 10.1103/PhysRevE.60.2366} {\bibfield  {journal} {\bibinfo
  {journal} {Phys. Rev. E}\ }\textbf {\bibinfo {volume} {60}},\ \bibinfo
  {pages} {2366--2376} (\bibinfo {year} {1999}{\natexlab{b}})}\BibitemShut
  {NoStop}%
\bibitem [{\citenamefont {Barras}\ \emph {et~al.}(2019)\citenamefont {Barras},
  \citenamefont {Aldam}, \citenamefont {Roch}, \citenamefont {Brener},
  \citenamefont {Bouchbinder},\ and\ \citenamefont {Molinari}}]{Barras2019}%
  \BibitemOpen
  \bibfield  {author} {\bibinfo {author} {\bibfnamefont {F.}~\bibnamefont
  {Barras}}, \bibinfo {author} {\bibfnamefont {M.}~\bibnamefont {Aldam}},
  \bibinfo {author} {\bibfnamefont {T.}~\bibnamefont {Roch}}, \bibinfo {author}
  {\bibfnamefont {E.~A.}\ \bibnamefont {Brener}}, \bibinfo {author}
  {\bibfnamefont {E.}~\bibnamefont {Bouchbinder}}, \ and\ \bibinfo {author}
  {\bibfnamefont {J.-F.}\ \bibnamefont {Molinari}},\ }\bibfield  {title}
  {\enquote {\bibinfo {title} {Emergence of cracklike behavior of frictional
  rupture: The origin of stress drops},}\ }\href {\doibase
  10.1103/PhysRevX.9.041043} {\bibfield  {journal} {\bibinfo  {journal} {Phys.
  Rev. X}\ }\textbf {\bibinfo {volume} {9}},\ \bibinfo {pages} {041043}
  (\bibinfo {year} {2019})}\BibitemShut {NoStop}%
\bibitem [{\citenamefont {Geubelle}\ and\ \citenamefont
  {Rice}(1995)}]{geubelle1995spectral}%
  \BibitemOpen
  \bibfield  {author} {\bibinfo {author} {\bibfnamefont {P.~H.}\ \bibnamefont
  {Geubelle}}\ and\ \bibinfo {author} {\bibfnamefont {J.~R.}\ \bibnamefont
  {Rice}},\ }\bibfield  {title} {\enquote {\bibinfo {title} {A spectral method
  for three-dimensional elastodynamic fracture problems},}\ }\href@noop {}
  {\bibfield  {journal} {\bibinfo  {journal} {Journal of the Mechanics and
  Physics of Solids}\ }\textbf {\bibinfo {volume} {43}},\ \bibinfo {pages}
  {1791--1824} (\bibinfo {year} {1995})}\BibitemShut {NoStop}%
\bibitem [{\citenamefont {Breitenfeld}\ and\ \citenamefont
  {Geubelle}(1998)}]{breitenfeld1998numerical}%
  \BibitemOpen
  \bibfield  {author} {\bibinfo {author} {\bibfnamefont {M.~S.}\ \bibnamefont
  {Breitenfeld}}\ and\ \bibinfo {author} {\bibfnamefont {P.~H.}\ \bibnamefont
  {Geubelle}},\ }\bibfield  {title} {\enquote {\bibinfo {title} {Numerical
  analysis of dynamic debonding under 2d in-plane and 3d loading},}\
  }\href@noop {} {\bibfield  {journal} {\bibinfo  {journal} {International
  Journal of Fracture}\ }\textbf {\bibinfo {volume} {93}},\ \bibinfo {pages}
  {13--38} (\bibinfo {year} {1998})}\BibitemShut {NoStop}%
\bibitem [{\citenamefont {Camacho}\ and\ \citenamefont
  {Ortiz}(1996)}]{camacho1996computational}%
  \BibitemOpen
  \bibfield  {author} {\bibinfo {author} {\bibfnamefont {G.~T.}\ \bibnamefont
  {Camacho}}\ and\ \bibinfo {author} {\bibfnamefont {M.}~\bibnamefont
  {Ortiz}},\ }\bibfield  {title} {\enquote {\bibinfo {title} {Computational
  modelling of impact damage in brittle materials},}\ }\href@noop {} {\bibfield
   {journal} {\bibinfo  {journal} {International Journal of Solids and
  Structures}\ }\textbf {\bibinfo {volume} {33}},\ \bibinfo {pages}
  {2899--2938} (\bibinfo {year} {1996})}\BibitemShut {NoStop}%
\bibitem [{\citenamefont {Andrews}\ and\ \citenamefont
  {Ben-Zion}(1997)}]{andrews1997wrinkle}%
  \BibitemOpen
  \bibfield  {author} {\bibinfo {author} {\bibfnamefont {D.~J.}\ \bibnamefont
  {Andrews}}\ and\ \bibinfo {author} {\bibfnamefont {Y.}~\bibnamefont
  {Ben-Zion}},\ }\bibfield  {title} {\enquote {\bibinfo {title} {Wrinkle-like
  slip pulse on a fault between different materials},}\ }\href@noop {}
  {\bibfield  {journal} {\bibinfo  {journal} {Journal of Geophysical Research:
  Solid Earth}\ }\textbf {\bibinfo {volume} {102}},\ \bibinfo {pages}
  {553--571} (\bibinfo {year} {1997})}\BibitemShut {NoStop}%
\bibitem [{\citenamefont {Ben-Zion}\ and\ \citenamefont
  {Huang}(2002)}]{ben2002dynamic}%
  \BibitemOpen
  \bibfield  {author} {\bibinfo {author} {\bibfnamefont {Y.}~\bibnamefont
  {Ben-Zion}}\ and\ \bibinfo {author} {\bibfnamefont {Y.}~\bibnamefont
  {Huang}},\ }\bibfield  {title} {\enquote {\bibinfo {title} {Dynamic rupture
  on an interface between a compliant fault zone layer and a stiffer
  surrounding solid},}\ }\href@noop {} {\bibfield  {journal} {\bibinfo
  {journal} {Journal of Geophysical Research: Solid Earth}\ }\textbf {\bibinfo
  {volume} {107}},\ \bibinfo {pages} {2042--2054} (\bibinfo {year}
  {2002})}\BibitemShut {NoStop}%
\bibitem [{\citenamefont {Rubin}\ and\ \citenamefont
  {Ampuero}(2007)}]{rubin2007aftershock}%
  \BibitemOpen
  \bibfield  {author} {\bibinfo {author} {\bibfnamefont {A.~M.}\ \bibnamefont
  {Rubin}}\ and\ \bibinfo {author} {\bibfnamefont {J.-P.}\ \bibnamefont
  {Ampuero}},\ }\bibfield  {title} {\enquote {\bibinfo {title} {Aftershock
  asymmetry on a bimaterial interface},}\ }\href@noop {} {\bibfield  {journal}
  {\bibinfo  {journal} {Journal of Geophysical Research: Solid Earth}\ }\textbf
  {\bibinfo {volume} {112}},\ \bibinfo {pages} {B05307} (\bibinfo {year}
  {2007})}\BibitemShut {NoStop}%
\bibitem [{\citenamefont {Ben-Zion}(2001)}]{ben2001dynamic}%
  \BibitemOpen
  \bibfield  {author} {\bibinfo {author} {\bibfnamefont {Y.}~\bibnamefont
  {Ben-Zion}},\ }\bibfield  {title} {\enquote {\bibinfo {title} {Dynamic
  ruptures in recent models of earthquake faults},}\ }\href@noop {} {\bibfield
  {journal} {\bibinfo  {journal} {Journal of the Mechanics and Physics of
  Solids}\ }\textbf {\bibinfo {volume} {49}},\ \bibinfo {pages} {2209--2244}
  (\bibinfo {year} {2001})}\BibitemShut {NoStop}%
\bibitem [{\citenamefont {Rice}\ \emph {et~al.}(2001)\citenamefont {Rice},
  \citenamefont {Lapusta},\ and\ \citenamefont {Ranjith}}]{rice2001rate}%
  \BibitemOpen
  \bibfield  {author} {\bibinfo {author} {\bibfnamefont {J.~R.}\ \bibnamefont
  {Rice}}, \bibinfo {author} {\bibfnamefont {N.}~\bibnamefont {Lapusta}}, \
  and\ \bibinfo {author} {\bibfnamefont {K.}~\bibnamefont {Ranjith}},\
  }\bibfield  {title} {\enquote {\bibinfo {title} {Rate and state dependent
  friction and the stability of sliding between elastically deformable
  solids},}\ }\href@noop {} {\bibfield  {journal} {\bibinfo  {journal} {Journal
  of the Mechanics and Physics of Solids}\ }\textbf {\bibinfo {volume} {49}},\
  \bibinfo {pages} {1865--1898} (\bibinfo {year} {2001})}\BibitemShut {NoStop}%
\bibitem [{\citenamefont {Rubinstein}\ \emph {et~al.}(2006)\citenamefont
  {Rubinstein}, \citenamefont {Shay}, \citenamefont {Cohen},\ and\
  \citenamefont {Fineberg}}]{Rubinstein2006}%
  \BibitemOpen
  \bibfield  {author} {\bibinfo {author} {\bibfnamefont {S.~M.}\ \bibnamefont
  {Rubinstein}}, \bibinfo {author} {\bibfnamefont {M.}~\bibnamefont {Shay}},
  \bibinfo {author} {\bibfnamefont {G.}~\bibnamefont {Cohen}}, \ and\ \bibinfo
  {author} {\bibfnamefont {J.}~\bibnamefont {Fineberg}},\ }\bibfield  {title}
  {\enquote {\bibinfo {title} {Crack-like processes governing the onset of
  frictional slip},}\ }\href {\doibase 10.1007/s10704-006-0049-8} {\bibfield
  {journal} {\bibinfo  {journal} {International Journal of Fracture}\ }\textbf
  {\bibinfo {volume} {140}},\ \bibinfo {pages} {201--212} (\bibinfo {year}
  {2006})}\BibitemShut {NoStop}%
\bibitem [{\citenamefont {Schallamach}(1971)}]{schallamach1971does}%
  \BibitemOpen
  \bibfield  {author} {\bibinfo {author} {\bibfnamefont {A.}~\bibnamefont
  {Schallamach}},\ }\bibfield  {title} {\enquote {\bibinfo {title} {How does
  rubber slide?}}\ }\href@noop {} {\bibfield  {journal} {\bibinfo  {journal}
  {Wear}\ }\textbf {\bibinfo {volume} {17}},\ \bibinfo {pages} {301--312}
  (\bibinfo {year} {1971})}\BibitemShut {NoStop}%
\bibitem [{\citenamefont {Baumberger}\ \emph {et~al.}(2002)\citenamefont
  {Baumberger}, \citenamefont {Caroli},\ and\ \citenamefont
  {Ronsin}}]{baumberger2002self}%
  \BibitemOpen
  \bibfield  {author} {\bibinfo {author} {\bibfnamefont {T.}~\bibnamefont
  {Baumberger}}, \bibinfo {author} {\bibfnamefont {C.}~\bibnamefont {Caroli}},
  \ and\ \bibinfo {author} {\bibfnamefont {O.}~\bibnamefont {Ronsin}},\
  }\bibfield  {title} {\enquote {\bibinfo {title} {Self-healing slip pulses
  along a gel/glass interface},}\ }\href@noop {} {\bibfield  {journal}
  {\bibinfo  {journal} {Physical Review Letters}\ }\textbf {\bibinfo {volume}
  {88}},\ \bibinfo {pages} {075509} (\bibinfo {year} {2002})}\BibitemShut
  {NoStop}%
\bibitem [{\citenamefont {Berman}\ \emph {et~al.}(2020)\citenamefont {Berman},
  \citenamefont {Cohen},\ and\ \citenamefont {Fineberg}}]{Berman_2020}%
  \BibitemOpen
  \bibfield  {author} {\bibinfo {author} {\bibfnamefont {N.}~\bibnamefont
  {Berman}}, \bibinfo {author} {\bibfnamefont {G.}~\bibnamefont {Cohen}}, \
  and\ \bibinfo {author} {\bibfnamefont {J.}~\bibnamefont {Fineberg}},\
  }\bibfield  {title} {\enquote {\bibinfo {title} {Dynamics and properties of
  the cohesive zone in rapid fracture and friction},}\ }\href@noop {}
  {\bibfield  {journal} {\bibinfo  {journal} {Physical Review Letters}\ }
  (\bibinfo {year} {2020})}\BibitemShut {NoStop}%
\bibitem [{\citenamefont {Adda-Bedia}\ and\ \citenamefont
  {Amar}(2003)}]{AddaBedia2003}%
  \BibitemOpen
  \bibfield  {author} {\bibinfo {author} {\bibfnamefont {M.}~\bibnamefont
  {Adda-Bedia}}\ and\ \bibinfo {author} {\bibfnamefont {M.~Ben}\ \bibnamefont
  {Amar}},\ }\bibfield  {title} {\enquote {\bibinfo {title} {Self-sustained
  slip pulses of finite size between dissimilar materials},}\ }\href {\doibase
  10.1016/S0022-5096(03)00068-1} {\bibfield  {journal} {\bibinfo  {journal}
  {Journal of the Mechanics and Physics of Solids}\ }\textbf {\bibinfo {volume}
  {51}},\ \bibinfo {pages} {1849--1861} (\bibinfo {year} {2003})}\BibitemShut
  {NoStop}%
\bibitem [{\citenamefont {Wang}\ \emph {et~al.}(1998)\citenamefont {Wang},
  \citenamefont {Huang}, \citenamefont {Rosakis},\ and\ \citenamefont
  {Liu}}]{wang1998effect}%
  \BibitemOpen
  \bibfield  {author} {\bibinfo {author} {\bibfnamefont {W.}~\bibnamefont
  {Wang}}, \bibinfo {author} {\bibfnamefont {Y.}~\bibnamefont {Huang}},
  \bibinfo {author} {\bibfnamefont {A.~J.}\ \bibnamefont {Rosakis}}, \ and\
  \bibinfo {author} {\bibfnamefont {C.}~\bibnamefont {Liu}},\ }\bibfield
  {title} {\enquote {\bibinfo {title} {Effect of elastic mismatch in intersonic
  crack propagation along a bimaterial interface},}\ }\href@noop {} {\bibfield
  {journal} {\bibinfo  {journal} {Engineering Fracture Mechanics}\ }\textbf
  {\bibinfo {volume} {61}},\ \bibinfo {pages} {471--485} (\bibinfo {year}
  {1998})}\BibitemShut {NoStop}%
\end{thebibliography}%

\begin{appendix}

\newpage

\section{The asymptotic elastic fields}
\label{app:fields}

In the framework of the problem defined by Eqs.~(\ref{eq:bulk}--\ref{eq:bc-tau}), the asymptotic stress field in the vicinity of the rupture front exhibits a universal square root singularity that is independent of rupture dynamics, bimaterial geometry and/or applied remote loading. This fundamental result has been found and verified using different methods~\cite{deng1993propagating,AddaBedia2003,shlomai_PNAS}. Among these, the so-called Williams expansion method is the most straightforward~\cite{shlomai_PNAS} one. In the upper half-plane corresponding to the soft material $(y>0)$, the square root terms of the different elastic field components are explicitly given by~\cite{shlomai_PNAS}
\bea
\partial_x u^{(1)}_{x}(x,y)&=& \frac{-\mu_1^{-1}K(t)}{2 a_1+(1+b_1^2)\alpha}\left\{\Im\left[\frac{1}{\sqrt{2\pi\zeta_{a1}}}\right]+b_1 \alpha\Im\left[\frac{1}{\sqrt{2\pi\zeta_{b1}}}\right] \right\}\;, \\
\partial_x u^{(1)}_{y}(x,y)&=& \frac{-\mu_1^{-1}K(t)}{2 a_1+(1+b_1^2)\alpha}\left\{ a_1\Re\left[\frac{1}{\sqrt{2\pi\zeta_{a1}}}\right] +\alpha \Re\left[\frac{1}{\sqrt{2\pi\zeta_{b1}}}\right] \right]\;, \\
\sigma^{(1)}_{xx}(x,y)&=&\frac{K(t)}{2 a_1+(1+b_1^2)\alpha}\left\{(1-b_1^2+2a_1^2) \Im\left[\frac{1}{\sqrt{2\pi\zeta_{a1}}}\right] +2b_1 \alpha\Im\left[\frac{1}{\sqrt{2\pi\zeta_{b1}}}\right]\right\}\; , \\
\sigma^{(1)}_{yy}(x,y)&=&\frac{-K(t)}{2 a_1+(1+b_1^2)\alpha}\left\{(1+b_1^2) \Im\left[\frac{1}{\sqrt{2\pi\zeta_{a1}}}\right]+2b_1 \alpha\Im\left[\frac{1}{\sqrt{2\pi\zeta_{b1}}}\right]\right\} \;, \\
\sigma^{(1)}_{xy}(x,y)&=&\frac{K(t)}{2 a_1+(1+b_1^2)\alpha}\left\{2 a_1\Re\left[\frac{1}{2\pi\sqrt{\zeta_{a1}}}\right]+(1+b_1^2)\alpha\Re\left[\frac{1}{\sqrt{2\pi\zeta_{b1}}}\right]\right\}\; ,
\eea
where $\zeta_{a1}=x-\ell(t)+ia_1y$ and $\zeta_{b1}=x-\ell(t)+ib_1y$ are complex coordinates. Here, $\Im$ (resp. $\Re$) denotes the imaginary (resp. real) part and $\alpha$ is a real velocity-dependent constant that is given by:
\beq
\alpha=-\frac{ (1 - b_2^2) (a_2 (1 + b_1^2)+  a_1   (1 + b_2^2)) \mu_1 +a_1D_2 (\mu_2-\mu_1)}{((1 + b_1^2) (1 + b_2^2-2a_2b_2) +2a_2 b_1(1-b_2^2)) \mu_1 + D_2 \mu_2}\; .
\label{eq:alpha}
\eeq
The functions $a_n$, $b_n$ and $D_n$ are defined in~Eq.(\ref{eq:speeds}), within the main text. Equivalently, in the lower half plane corresponding to the stiff material $(y<0)$, one has
\bea
\partial_x u^{(2)}_{x}(x,y)&=& \frac{-\mu_2^{-1}K(t)}{2 a_1+(1+b_1^2)\alpha}\left\{\beta\Im\left[\frac{1}{\sqrt{2\pi\zeta_{a2}}}\right]+ b_2\gamma \Im\left[\frac{1}{\sqrt{2\pi\zeta_{b2}}}\right] \right\}\;, \\
\partial_x u^{(1)}_{y}(x,y)&=& \frac{-\mu_2^{-1}K(t)}{2 a_1+(1+b_1^2)\alpha}\left\{ a_2\beta\Re\left[\frac{1}{\sqrt{2\pi\zeta_{a2}}}\right] +\gamma \Re\left[\frac{1}{\sqrt{2\pi\zeta_{b2}}}\right] \right]\;, \\
\sigma^{(2)}_{xx}(x,y)&=&\frac{K(t)}{2 a_1+(1+b_1^2)\alpha}\left\{(1-b_2^2+2a_2^2) \beta \Im\left[\frac{1}{\sqrt{2\pi\zeta_{a2}}}\right] +2b_2 \gamma \Im\left[\frac{1}{\sqrt{2\pi\zeta_{b2}}}\right]\right\}\; , \\
\sigma^{(2)}_{yy}(x,y)&=&\frac{-K(t)}{ 2 a_1+(1+b_1^2)\alpha}\left\{(1+b_2^2) \beta \Im\left[\frac{1}{\sqrt{2\pi\zeta_{a2}}}\right]+2b_2 \gamma\Im\left[\frac{1}{\sqrt{2\pi\zeta_{b2}}}\right]\right\} \; , \\
\sigma^{(2)}_{xy}(x,y)&=&\frac{K(t)}{2 a_1+(1+b_1^2)\alpha}\left\{2 a_2\beta\Re\left[\frac{1}{\sqrt{2\pi\zeta_{a2}}}\right]+(1+b_2^2)\gamma\Re\left[\frac{1}{\sqrt{2\pi\zeta_{b2}}}\right]\right\}\ .
\eea
where $\zeta_{a2}=x-\ell(t)+ia_2y$ and $\zeta_{b2}=x-\ell(t)+ib_2y$ are the corresponding complex coordinates. The  real velocity dependant constants $\beta$ and $\gamma$ are given by
\bea
\beta&=&\frac{  ((1 + b_2^2) (1 + b_1^2-2a_1b_1) +2a_1 b_2(1-b_1^2))  \mu_2+D_1 \mu_1}{((1 + b_1^2) (1 + b_2^2-2a_2b_2) +2a_2 b_1(1-b_2^2)) \mu_1 + D_2 \mu_2}
\; , \label{eq:beta}\\
\gamma&=&-\frac{ (1-b_1^2)(a_1(1+b_2^2)+a_2(1+b_1^2)) \mu_2+a_2D_1(\mu_1-\mu_2)}{((1 + b_1^2) (1 + b_2^2-2a_2b_2) +2a_2 b_1(1-b_2^2)) \mu_1 + D_2 \mu_2}
\;.\label{eq:gamma}
\eea
They satisfy the identities
\bea
(1+b_1^2)+2 b_1\alpha &=&-(1+b_2^2)\beta-2b_2\gamma\;,\\
2 a_1+(1+b_1^2)\alpha&=&2a_2\beta+(1+b_2^2)\gamma\;,\\
\mu_2(a_1+\alpha)&=&\mu_1(a_2\beta+\gamma)\;.
\eea
Notice that the parameters $\alpha(c)$, $\beta(c)$ and $\gamma(c)$ are not physically relevant as such but rather specific algebraic combinations of them. Following Weertman~\cite{weertman1980unstable}, we define the velocity dependant parameters $\mu^*(c)$ and $\overline{\mu}(c)$ as follow
\bea
\mu^*(c)&=&\frac{1+b_1^2+2b_1\alpha}{\mu_1^{-1}(1+b_1\alpha)+\mu_2^{-1}(\beta+b_2\gamma)}\;,\\
\overline{\mu}(c)&=&\frac{2a_1+(1+b_1^2)\alpha}{\mu_1^{-1}(1+b_1\alpha)+\mu_2^{-1}(\beta+b_2\gamma)}\;.
\label{eq:weertmandef}
\eea
In addition, the Weertman function introduced in Sec.~\ref{sec:fields} is defined by
\beq
W(c) =\frac{\mu^*(c)}{\overline{\mu}(c)} \;.
\eeq
Simple algebraic manipulations using Eqs.~(\ref{eq:alpha},\ref{eq:beta},\ref{eq:gamma}) allow us to write $W(c)$ under the form given by Eq.~(\ref{eq:weertman}).

Eqs.~(\ref{eq:sxy},\ref{eq:syy})  can be easily retrieved from the general expressions of the elastic field components by implementing the special case $y=0$. Moreover, one can show that
\beq
\sigma^{(n)}_{yy}(x,0)=\mu^*(c)\left[\partial_x u^{(1)}_{x}(x,0^+)-\partial_x u^{(2)}_{x}(x,0^-)\right]\,.
\label{eq:coupling}
\eeq
Specifically, the parameter $\mu^*(c)$ and consequently the Weertman function $W(c)$ highlight induced bimaterial coupling between normal loading and slip along the frictional interface.

It is noteworthy that the asymptotic stress and strain fields involve the motion of rupture front only through the instantaneous crack tip speed $c(t)=\dot{\ell}(t)$. Moreover, all information about the loading conditions and bimaterial geometric scales are embedded in the dynamic stress intensity factor $K(t)$. An immediate consequence is that the near-tip fields for nonuniform motion are identical to those describing steady state rupture propagation. In particular, the square root singular behavior of the asymptotic fields exhibit universal angular variations, in the sense that they depend only on the instantaneous rupture tip speed and bimaterial elastic parameters.

Finally, the real nature of the singularity is closely related to the imposed debonding behavior~\cite{shlomai_PNAS}. For a rupture propagating along a bimaterial interface, the condition for which the corresponding asymptotic stress field exhibits a complex or a real singularity can be rationalized as follows. If interfacial contact immediately behind the rupture front is lost, the stress field's asymptotic behavior obeys $\sigma\sim \Re[K/r^\lambda]$, where $\lambda$ is a complex exponent and $K$ is a complex stress intensity-like factor~\cite{Rice1988Elastic,yang1991mechanics}. If, however, contact between the two dissimilar bodies is {\it preserved}, the stress field singularity in the vicinity of the rupture tip becomes  $\sigma\sim K/r^q$, where $0\leq q\leq 1/2$ is a positive real exponent and $K$ is a real stress intensity-like factor. The real nature of the exponent is independent of the particular choice of the friction law, but its value depends on it~\cite{AddaBedia2003}. Moreover, these features are independent of the rupture front speed. They hold for all modes of rupture front propagation; these include quasistatic, subsonic, transonic and supershear dynamics~\cite{deng1993propagating,wang1998effect,shlomai_PNAS,shlomai_jgr}.

\section{The energy release rates}
\label{app:ERR}

For a crack propagating in a homogenous material, it is known that the energy release rate is path independent so long as the contour of integration remains in the near tip region, where the singular terms of the elastic fields are dominant~\cite{freund1998dynamic,adda1999dynamic,Adda1999}. It is straightforward to generalize this result to our current problem for both $G_n$. Without any loss of generality, this property enables us to choose convenient contours over which the integration can be performed straightforwardly.  Following~\cite{freund1998dynamic}, we will use for ${\mathcal C}_n$ rectangular contours whose size $2\eta_x\times\eta_y$ is shrunk onto the rupture tip  (see Fig.~\ref{fig:problem2}). This is a convenient choice, as the computation of the energy release rate contributions of both soft and stiff materials involves only integrals along segments parallel to the $x$-axis. Equation~(\ref{eq:Gsep}) is then simplified into~\cite{freund1998dynamic}
\bea
G_1&=&-\lim_{\eta_x\rightarrow 0}\left\{\lim_{\eta_y\rightarrow 0}\int_{-\eta_x+\ell(t)}^{\eta_x+\ell(t)}\sigma^{(1)}_{iy}(x,\eta_y)\frac{\partial u^{(1)}_i(x,\eta_y)}{\partial x}dx\right\}\;,
\label{eq:intG1}\\
G_2&=&\lim_{\eta_x\rightarrow 0}\left\{\lim_{\eta_y\rightarrow 0}\int_{-\eta_x+\ell(t)}^{\eta_x+\ell(t)}\sigma^{(2)}_{iy}(x,-\eta_y)\frac{\partial u^{(2)}_i(x,-\eta_y)}{\partial x}dx\right\}\;,
\label{eq:intG2}
\eea
where we have used $\partial u^{(n)}_i/\partial t=-c\partial u^{(n)}_i/\partial x$, which is always satisfied near the rupture front. Using the expressions of the asymptotic stress and strain fields given in Appendix~\ref{app:fields}, one can show that
\bea
&&\sigma^{(1)}_{iy}(x,\eta_y)\partial_x u^{(1)}_i(x,\eta_y)=\frac{(1-b_1^2)K^2(t)}{4\pi\mu_1\left(2a_1+(1+b_1^2)\alpha\right)^2}\left[ \frac{a^2_1 \eta_y}{(x-\ell(t))^2+a_1^2\eta_y^2}- \frac{b^2_1\alpha^2 \eta_y}{(x-\ell(t))^2+b_1^2\eta_y^2}\right] \ , \\
&&\sigma^{(2)}_{iy}(x,-\eta_y)\partial_x u^{(2)}_i(x,-\eta_y)= \frac{-(1-b_2^2)K^2(t)}{4\pi\mu_2\left(2a_1+(1+b_1^2)\alpha\right)^2}\left[ \frac{a^2_2\beta^2 \eta_y}{(x-\ell(t))^2+a_2^2\eta_y^2}- \frac{b^2_2\gamma^2 \eta_y}{(x-\ell(t))^2+b_2^2\eta_y^2}\right] \ .
\eea
The integrals in Eqs.~(\ref{eq:intG1},\ref{eq:intG2}) then yield
\bea
&&G_1=\frac{(1-b_1^2)(a_1 - b_1\alpha^2)}{4\mu_1\left(2a_1+(1+b_1^2)\alpha\right)^2}\,K^2(t)  \ , \\
&&G_2= \frac{(1-b_2^2)(a_2\beta^2 - b_2\gamma^2)}{4\mu_2\left(2a_1+(1+b_1^2)\alpha\right)^2}\,K^2(t) \ .
\eea
Finally, using the definitions of the constants $\alpha$, $\beta$ and $\gamma$ as given by Eqs.~(\ref{eq:alpha},\ref{eq:beta},\ref{eq:gamma}), simple algebraic manipulations allow us to write the energy release rate $G_n$ with the form given by Eq.~(\ref{eq:finalGn}).

\end{appendix}

\end{document}